\documentclass[review]{elsarticle}
 \biboptions{comma,sort&compress}

\usepackage[table]{xcolor}

\usepackage{graphicx}
\usepackage{amsmath}
\usepackage{here}
\usepackage{cuted}

\usepackage{here}
\usepackage[dvips]{epsfig}
\definecolor{bluette}{rgb}{.2,.4,0}
\definecolor{salmon}{rgb}{.9,0.68,0.5}
\definecolor{motive}{rgb}{0.2,1,.5}
\definecolor{list}{rgb}{0.3,.8,.1}
\definecolor{moe}{rgb}{1,.7,.5}
\definecolor{mote}{rgb}{.7,.5,.6}
\definecolor{pisello}{rgb}{.1,1,0}
\definecolor{orange}{rgb}{1,.7,0}
\definecolor{oliva}{rgb}{.1,.5,0.3}
\definecolor{greenda}{rgb}{0,.3,.2}
\definecolor{greenli}{rgb}{0.5,.8,.0}
\definecolor{blueda}{rgb}{0,.1,.6}
\definecolor{purple}{rgb}{.7,.1,.2}
\definecolor{marrone}{rgb}{1,0.7,0}
\definecolor{pinky}{rgb}{1,0.8,0.8}
\definecolor{rose}{rgb}{1,0.4,0}

\def\oliva{\color{oliva}}

\def\beq{\begin{equation}}
\def\eeq{\end{equation}}
\def\bea{\begin{eqnarray}}
\def\eea{\end{eqnarray}}
\def\bq{\begin{quote}}
\def\eq{\end{quote}}

\def\nnb{\nonumber}
\def\ga{\left(}
\def\dr{\right)}

\def\lrar{\Longrightarrow}
\def\lrar2{\longrightarrow}
																
\def\nnb{\nonumber}

\def\la{\langle}
\def\ra{\rangle}

\def\ba{\vspace*{-0.2cm}\begin{array}}
\def\ea{\end{array}\vspace*{-0.2cm}}

\def\b{$\bullet~$}
\def\d{$\diamond~$}

\def\als{\alpha_s}

\def\gg2{\la\alpha_s G^2 \ra}
\def\gg3{g^3f_{abc}\la G^aG^bG^c \ra}
\def\ggg4{\la\als^2G^4\ra}

\def\gg{\lag g^{2}_{s} G^2 \rag}
\def\ggg{\lag g^{3}_{s}G^3\rag}

\usepackage{amsmath}
\usepackage{slashed}
\usepackage{color}

\begin{document}
\begin{frontmatter}

\title{QCD condensates and $\alpha_s$  from $\tau$-decay}
\author{Stephan Narison
}
\address{Laboratoire
Univers et Particules de Montpellier (LUPM), CNRS-IN2P3, \\
Case 070, Place Eug\`ene
Bataillon, 34095 - Montpellier, France\\
and\\
Institute of High-Energy Physics of Madagascar (iHEPMAD)\\
University of Ankatso, Antananarivo 101, Madagascar}
\ead{snarison@yahoo.fr}


\date{\today}
\begin{abstract}
We improve the determinations of the QCD condensates within the SVZ expansion in the axial-vector (A) channel using the ratio of Laplace sum rule (LSR) ${\cal R}_{10}^A(\tau)$ within stability criteria and $\tau$-like higher moments ${\cal R}_{n,A}$ within stability for arbitrary $\tau$-mass squared $s_0$. We find the same violation of the factorization by a factor 6 of the four-quark condensate as  from $e^+e^-\to$ Hadrons data.  One can notice a systematic alternate sign and no exponential growth of the size of these condensates.   Then, we extract $\alpha_s$ from the lowest $\tau$-decay like moment.  We obtain to order $\alpha_s^4$ the conservative value from the $s_0$-stability  until  $M_\tau^2$ : $\alpha_s(M_\tau)\vert_A=0.3178(66)$ (FO) and 0.3380 (44) (CI) leading to\,: $\alpha_s(M_Z)\vert_A=0.1182(8)_{fit}(3)_{evol.}$ (FO) and  $0.1206(5)_{fit}(3)_{evol.}$\,(CI).  We extend the analysis to the  channel and find: $ \alpha_s(M_\tau)\vert_{V-A}=0.3135(83)$ (FO) and 0.3322 (81) (CI) leading to\,: $\alpha_s(M_Z)\vert_{V-A}=0.1177(10)_{fit}(3)_{evol.}$ (FO) and  $0.1200(9)_{fit}(3)_{evol.}$\,(CI). We observe  that in different channels ($e^+e^-\to$ Hadrons,\, A,\,V,\,V--A), the extraction of $\alpha_s(M_\tau)$ at the observed $\tau$-mass leads to an overestimate of its value.  Our determinations from these different channels lead to the mean\,: $ \alpha_s(M_\tau)=0.3140(44)$ (FO) and 0.3346 (35) (CI) leading to\,: $\alpha_s(M_Z)=0.1178(6)_{fit}(3)_{evol.}$ (FO) and  $0.1202(4)_{fit}(3)_{evol.}$\,(CI). Comparisons with some other results are done.

\end{abstract}
\end{frontmatter}
\newpage
\section{Introduction}
\vspace*{-0.2cm}

In this paper,  we pursue the determinations of the QCD condensates and $\alpha_s$ done in \cite{SNe,SNe2} for the $e^+e^-\to$  Hadrons and the vector (V) current
to the case of axial-vector (A) and V--A currents.  In so doing, we shall use the $\tau$ sum rule variable stability criteria
for the Laplace sum rule and the $M_\tau$ stability for the $\tau$-like moment sum rule.

Definitions and normalizations of observables will be the same as in Ref.\,\cite{SNe,SNe2} and will not be extensively discussed here.

\section{The axial-vector (A)  two-point function}
\subsection*{\b The two-point function}
We shall be concerned with the two-point correlator :
\bea
\hspace*{-0.6cm} 
\Pi^{\mu\nu}_{V(A)}(q^2)&=&i\hspace*{-0.1cm}\int \hspace*{-0.15cm}d^4x ~e^{-iqx}\la 0\vert {\cal T} {J^\mu_{V(A)}}(x)\ga {J^\nu_{V(A)}}(0)\dr^\dagger \vert 0\ra \nnb\\
&=&-(g^{\mu\nu}q^2-q^\mu q^\nu)\Pi^{(1)}_{V(A)}(q^2)+q^\mu q^\nu \Pi^{(0)}_{V(A)}(q^2)
 \label{eq:2-point}
 \eea
built from the T-product of the bilinear axial-vector current of $u,d$ quark fields:
\beq
 J^\mu_{V(A)}(x)=: \bar\psi_u\gamma^\mu(\gamma_5)\psi_d:.
\eeq
The upper  indices (0) and (1) correspond to the spin of the associated hadrons. The two-point function obeys the dispersion relation:
\beq
\Pi_{V(A)}(q^2)=\int_{t>}^\infty \frac{dt}{t-q^2-i\epsilon} \frac{1}{\pi}\,{\rm Im} \Pi_{V(A)}(t)+\cdots,
\eeq
where $\cdots$ are subtraction constants polynomial in $q^2$ and $t>$   is the hadronic threshold.
\subsection*{\b  QCD expression of the two-point function}
Within the SVZ-expansion\,\cite{SVZ}, the two-point function can be expressed in terms of the sum of higher and higher quark and gluon condensates:
\beq
4\pi^2\Pi_H(-Q^2,m_q^2,\mu)=\sum_{D=0,2,4,..}\hspace*{-0.25cm}\frac{C_{D,H}(Q^2,m_q^2,\mu)\la O_{D,H}(\mu)\ra}{(Q^2)^{D/2}}\equiv \sum_{D=0,2,4,..}\hspace*{-0.25cm}\frac {d_{D,H}} {(Q^2)^{D/2}}~, 
\label{eq:ope}
\eeq
where $H\equiv V(A)$, $\mu$ is the subtraction scale which separates the long (condensates) and short (Wilson coefficients) distance dynamics and $m_q$ is the quark mass. 

\d In the chiral limit\,\cite{BNP,BNP2}:
\bea
d_{2,A}&=&d_{2,V} = 0,\nnb\\
d_{4,A} &=&  d_{4,V} = \frac{\pi}{3}\la\alpha_s G^2\ra \ga 1+\frac{7}{6}a_s\dr 
,\nnb\\
d_{6,A} &=&  -\ga\frac{11}{7}\dr d_{6,V} =\frac{1408}{81}\pi^3\rho\alpha_s\la\bar \psi\psi\ra^2,\nnb\\
d_{8,A} &\approx&d_{8,V}\approx -\frac{39}{162}\pi \la\alpha_s G^2\ra^2
\label{eq:d8}
\eea
where $\rho$ measures the deviation from the factorization of the 4-quark condensates and the last equality for $d_8$  is based on vacuum saturation estimate of some classes of  computed dimension-8 diagrams. 

\d The perturbative expression of the spectral function is known to order $\alpha_s^4$\,\cite{LARIN,CHET4}. It reads for 3 flavours:
\beq
4{\pi}\,{\rm Im}\Pi_H(t)=  1+a_s+1.6398a_s^2-10.2839a_s^3-106.8798a_s^4+{\cal O}(a_s^5),
\label{eq:pt}
\eeq
where : 
\beq
a_s\equiv \frac{\alpha_s}{\pi}=\frac{2}{-\beta_1{\rm Log} (t/\Lambda^2)}+\cdots,
\eeq 
where $-\beta_1=(1/2)(11-2n_f/3)$ is the first coefficient of the $\beta$-function and $n_f$ is the number of quark flavours;  $\cdots$ stands for higher order terms which can e.g. be found in \cite{SNB2}. We shall use the value $\Lambda =(342\pm 8)$ MeV  for  $n_f=3$\:\, from the PDG average\,\cite{PDG}.

\subsection*{\b Spectral function from the data}
\begin{figure}[H]
\begin{center}
\includegraphics[width=10cm]{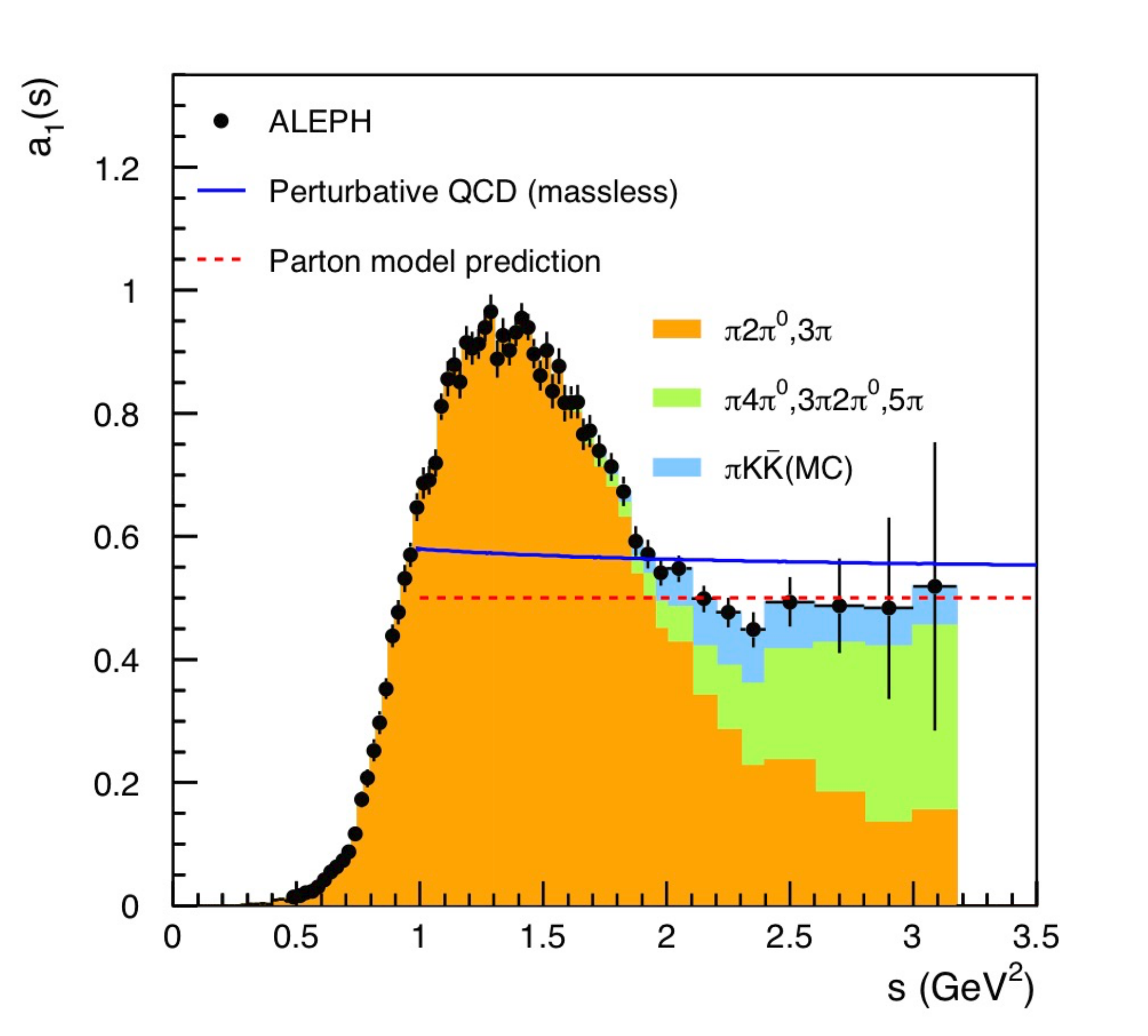}
\caption{\footnotesize  ALEPH data of the axial-vector spectral function. } \label{fig:aleph}
\end{center}
\vspace*{-0.5cm}
\end{figure}

We shall use the recent ALEPH data\,\cite{ALEPH}  in Fig.\,\ref{fig:aleph} for the spectral function $a_1(s)$.

\begin{figure}[H]
\begin{center}
\includegraphics[width=6.7cm]{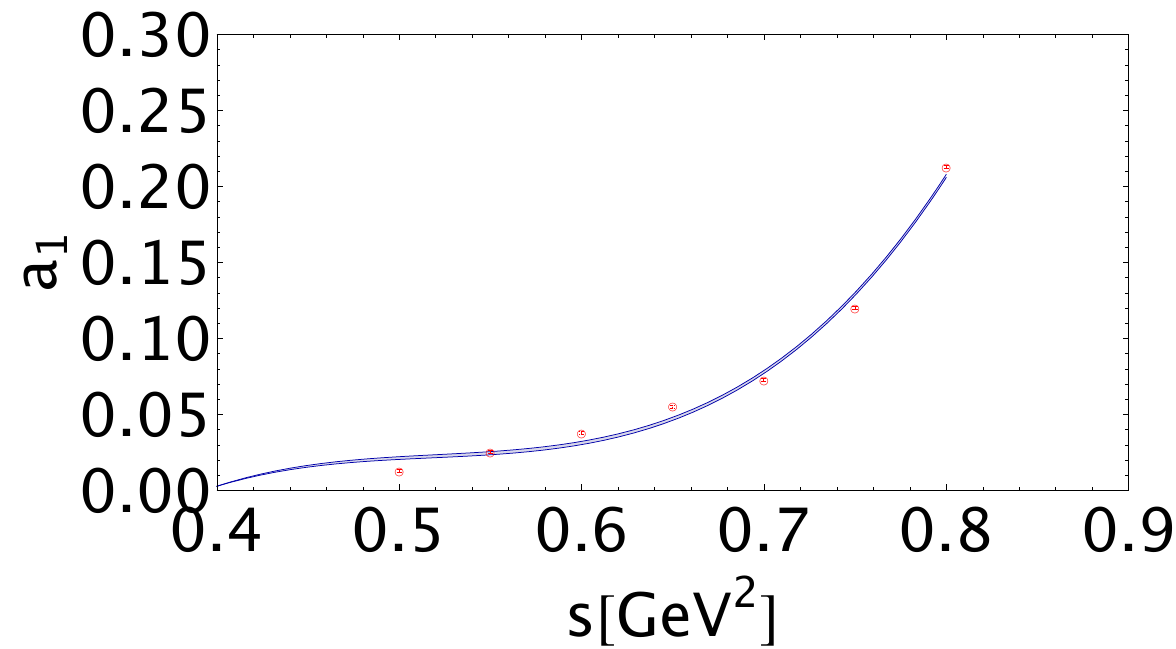}
\includegraphics[width=4.5cm]{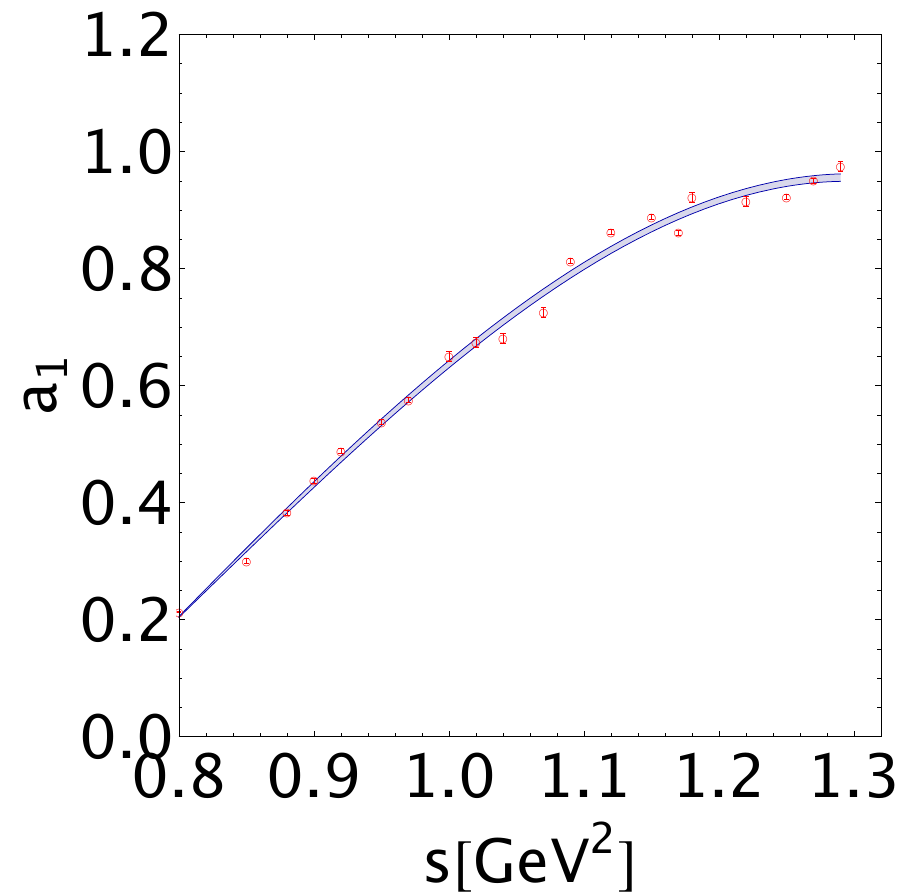}
\includegraphics[width=4.9cm]{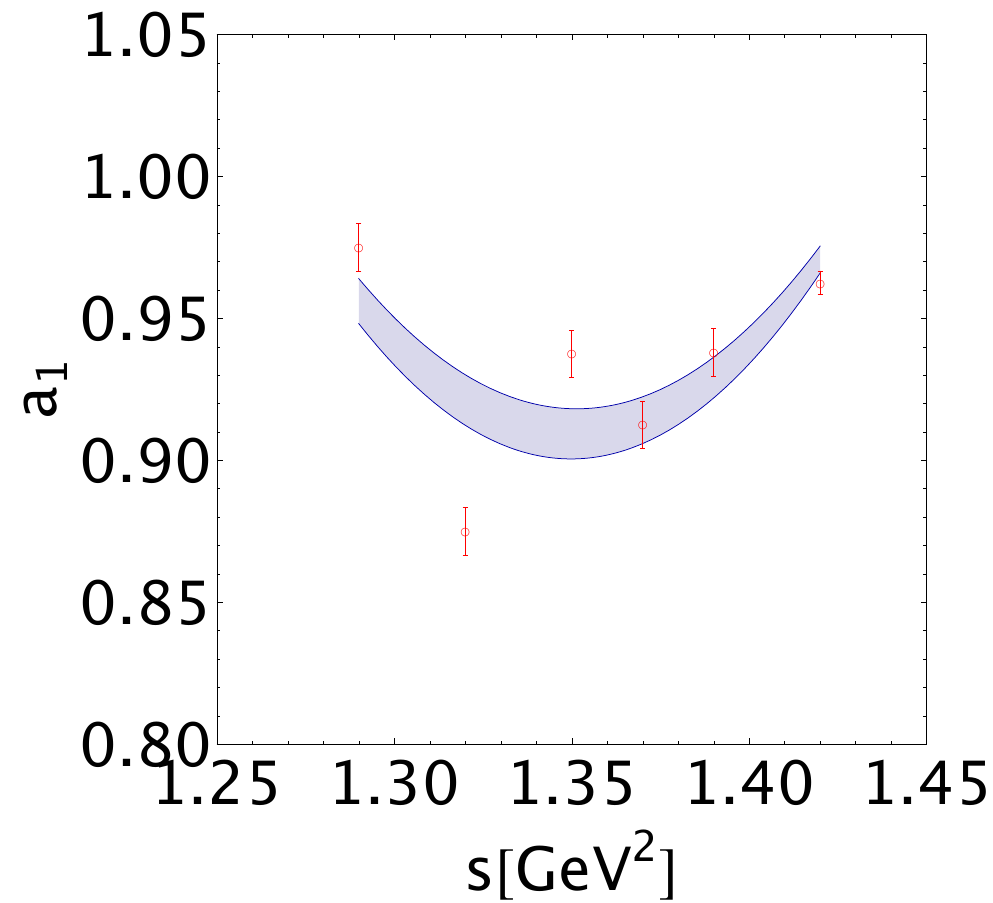}
\includegraphics[width=4.5cm]{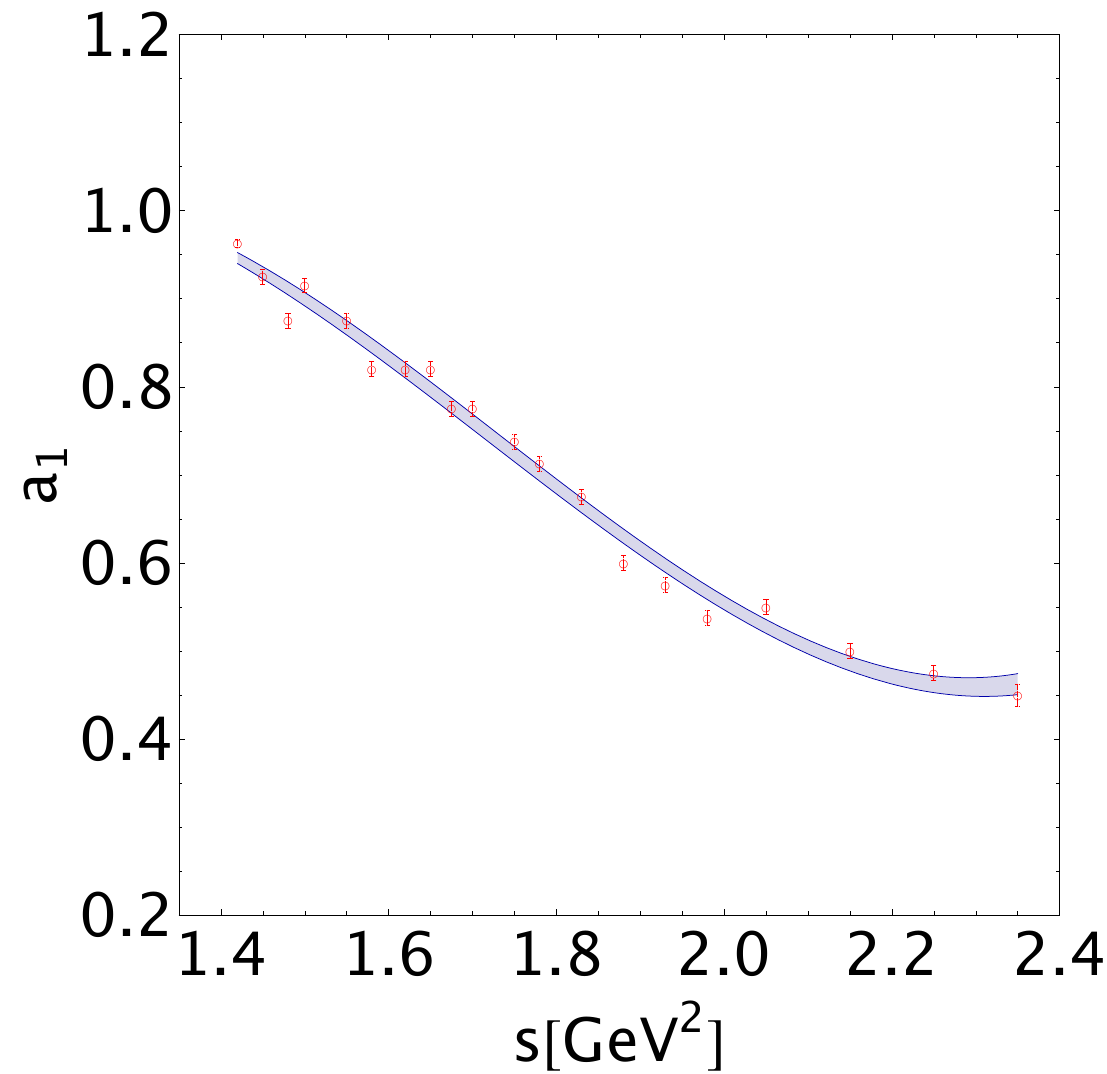}
\includegraphics[width=7.5cm]{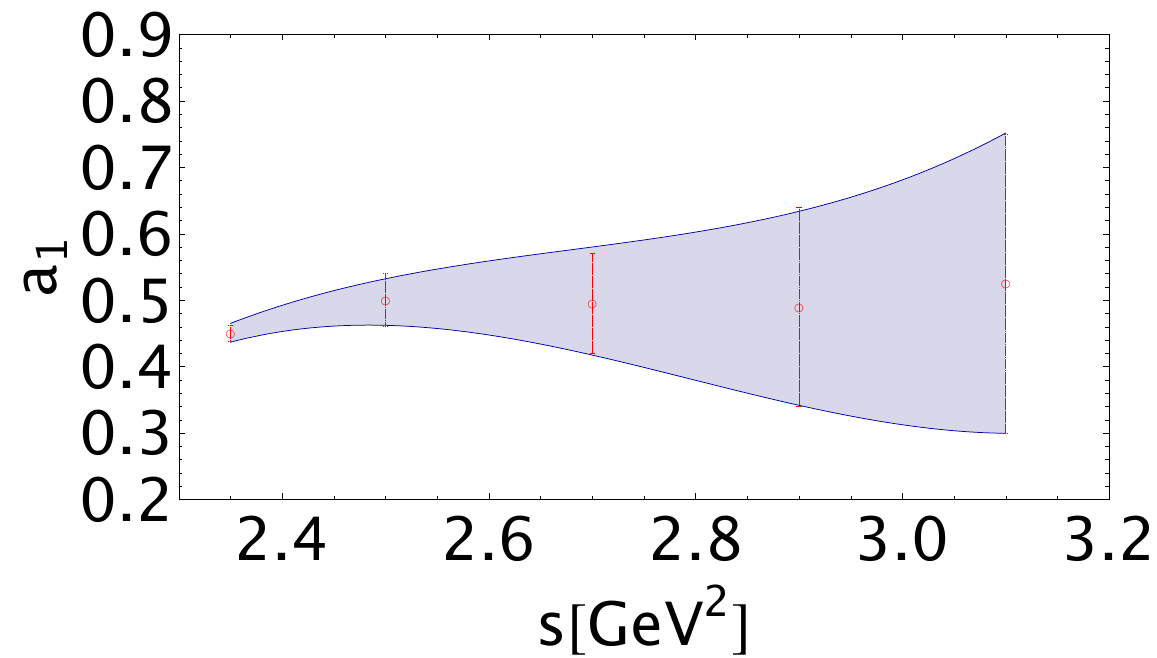}
\caption{\footnotesize  Fit of the data using a 2nd and 3rd order polynomial fits. } \label{fig:a1}
\end{center}
\vspace*{-0.5cm}
\end{figure}

Like in the case of the vector spectral function from $e^+e^-\to$ Hadrons\,\cite{SNe2}, we subdivide the region from $3\pi$ threshold to $M_\tau^2=3.16$ GeV$^2$ into different subregions in $s$ (units in  [GeV$^2$]):
\beq
s=[0.4,0.8],~[0.8,1.29],~[1.29,1.42],~[1.42,2.35], ~[2.35, 3.16],
\eeq
and fit the data with 3rd order polynomials using the optimized Mathematica program FindFit except for $[1.29,1.42]$ where a 2nd order polynomial is used.  
We show the different fits in the Fig.\,\ref{fig:a1}.
\section{The ratio of Laplace sum rule (LSR) moments}
Like in the case of vector channel, we shall use here the ratio of LSR moments\,\cite{SVZ,SNR,BELL}\footnote{For a recent review, see e.g.\,\cite{SNLSR}.}:
\beq
 {\cal R}^A_{10}(\tau)\equiv\frac{{\cal L}^c_{1}} {{\cal L}^c_0}= \frac{\int_{t>}^{t_c}dt~e^{-t\tau}t\, \frac{1}{\pi}\,{\rm Im}
 \Pi_H(t,\mu^2,m_q^2)}   {\int_{t>}^{t_c}dt~e^{-t\tau} \, \frac{1}{\pi}\,{\rm Im}
 \Pi_H(t,\mu^2,m_q^2)},
\label{eq:lsr}
\eeq
where $\tau$ is the LSR variable, $t>$   is the hadronic threshold.  Here $t_c$ is  the threshold of the ``QCD continuum" which parametrizes, from the discontinuity of the Feynman diagrams, the spectral function  ${\rm Im}\,\Pi_H(t,m_q^2,\mu^2)$.  $m_q$ is the quark mass and $\mu$ is an arbitrary subtraction point. 
\subsection*{\b QCD expression of the LSR moments}
To order $\alpha_s^4$, the perturbative (PT) expression of the lowest moment reads\,\cite{SNe}:
\beq
{\cal L}^{PT}_0(\tau)= \frac{3}{2}\tau^{-1}\Big{[} 1+a_s+2.93856\,a_s^2+ 6.2985\,a_s^3 + 22.2233\,a_s^4\Big{]}.
\eeq
Then, taking its derivative in $\tau$, one gets ${\cal L}_1(\tau)$ and then their ratio ${\cal R}_{10}(\tau)$. 

From Eq.\,\ref{eq:ope}, one can deduce the non-perturbative contribution to the lowest LSR moment\,:
\beq
{\cal L}^{NPT}_0(\tau) = \frac{3}{2}\tau^{-1}\sum_D \frac{d_D}{(D/2-1)!} \tau^{D/2} ~,
\label{eq:svz}
\eeq
from which one can deduce ${\cal L}^{NPT}_1$ and ${\cal R}^A_{10}$. 

\subsection*{\b   $d_{6,A}$ and $d_{8,A}$ from $ {\cal R}^A_{10}$}
\begin{figure}[hbt]
\begin{center}
\hspace*{0.cm}{\bf a)} \hspace*{8cm}{\bf b)}\\
\includegraphics[width=7.7cm]{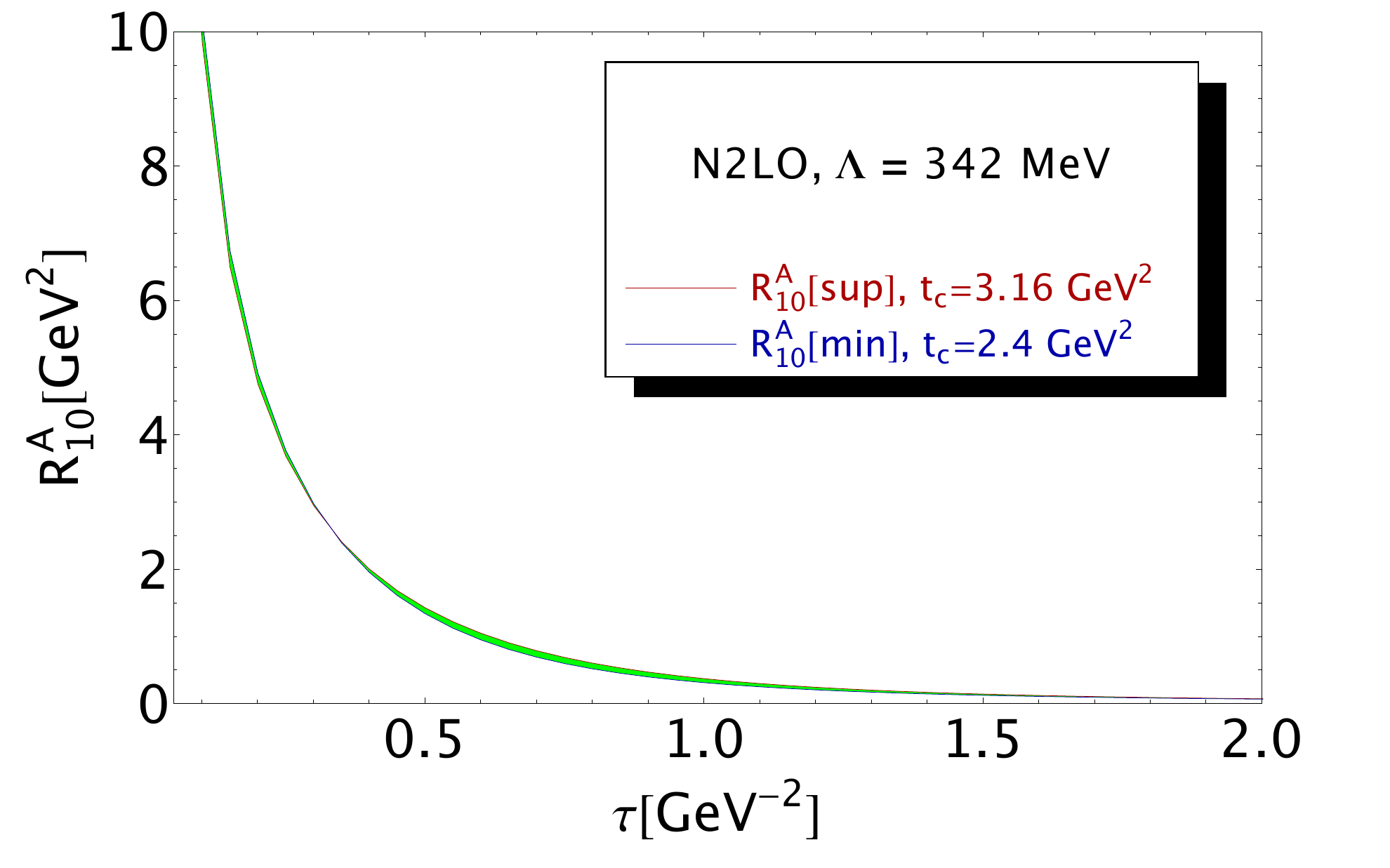}
\includegraphics[width=8.6cm]{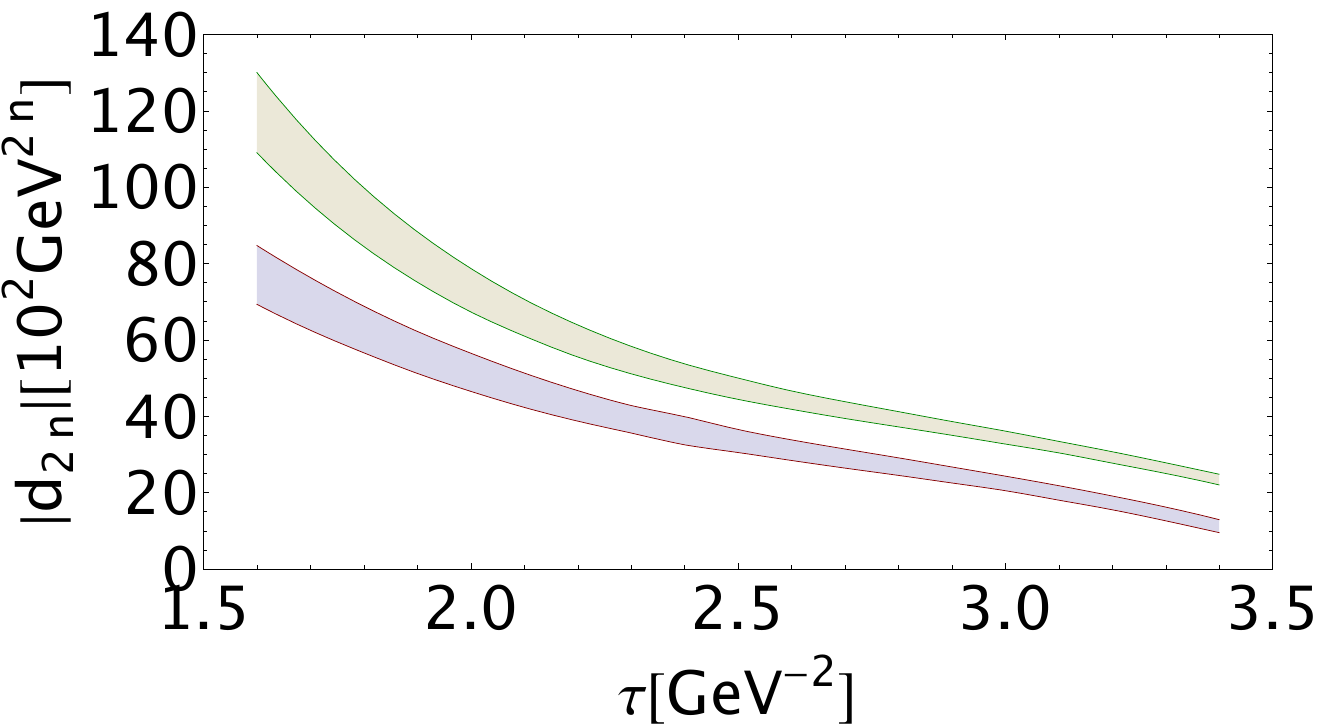}
\caption{\footnotesize {\bf a)}: $R^A_{10}$ versus the LSR variable $\tau$;  {\bf b)}:  $d_{6,A}$ and $d_{8,A}$ from the $ {\cal R}^A_{10}$} \label{fig:r10}
\end{center}
\vspace*{-0.5cm}
\end{figure}

We show in Fig.\,\ref{fig:r10}a) the $\tau$ behaviour of the phenomenological side of $ {\cal R}^A_{10}$ (experiment $\oplus$ QCD continuum beyond the threshold $t_c$) for values of $t_c$ around the physical $\tau$-lepton mass squared where the effect of $t_c$ is negligible. 

For determining $d_{6,A}$ and $d_{8,A}$, we use as input the more precise value of  $\la \alpha_s G^2\ra$  from the heavy quark mass-splittings and some other sum rules\,\cite{SNparam,SNcb1}:
\beq
\la\alpha_s G^2\ra =  (6.39\pm 0.35)\times 10^{-2}\,{\rm GeV^4},
\label{eq:g2}
\eeq
and perform a two-parameter fit $(d_{6,A},d_{8,A})$ by confronting the phenomenological and QCD side of $R^A_{10}$ for different values of $\tau$. The QCD coupling $\alpha_s$ is evaluated at the LSR sum rule scale $\tau$ where we use $\Lambda =(342\pm 8)$ MeV for $n_f=3$ flavours deduced from the PDG world average\,\cite{PDG}.
The results of the analysis are shown in  Fig.\,\ref{fig:r10}.  There is an inflexion point around $\tau\simeq 2.5 $ GeV$^{-2}$ at which we extract the optimal values of the condensates.  One should note that at this scale the OPE converges quite well in the vector channel\,\cite{SNe}. Then, we may expect that for the axial-vector the same feature occurs. We obtain the optimal result:
\beq
d_{6,A} =  (33.5\pm 3.0\pm 2.7)\times 10^{-2}\,{\rm GeV^6} ~ ~~~~~~~~~~~~~~ d_{8,A} = -(47.2\pm 2.8\pm 3.2)\times 10^{-2}\,{\rm GeV^8}
\label{eq:d68-ratio}
\eeq
where the 1st error comes from the fitting procedure and the 2nd one from the localization of $\tau\simeq  (2.5\pm 0.1)$ GeV$^{-2}$. 
\begin{figure}[hbt]
\begin{center}
\includegraphics[width=10cm]{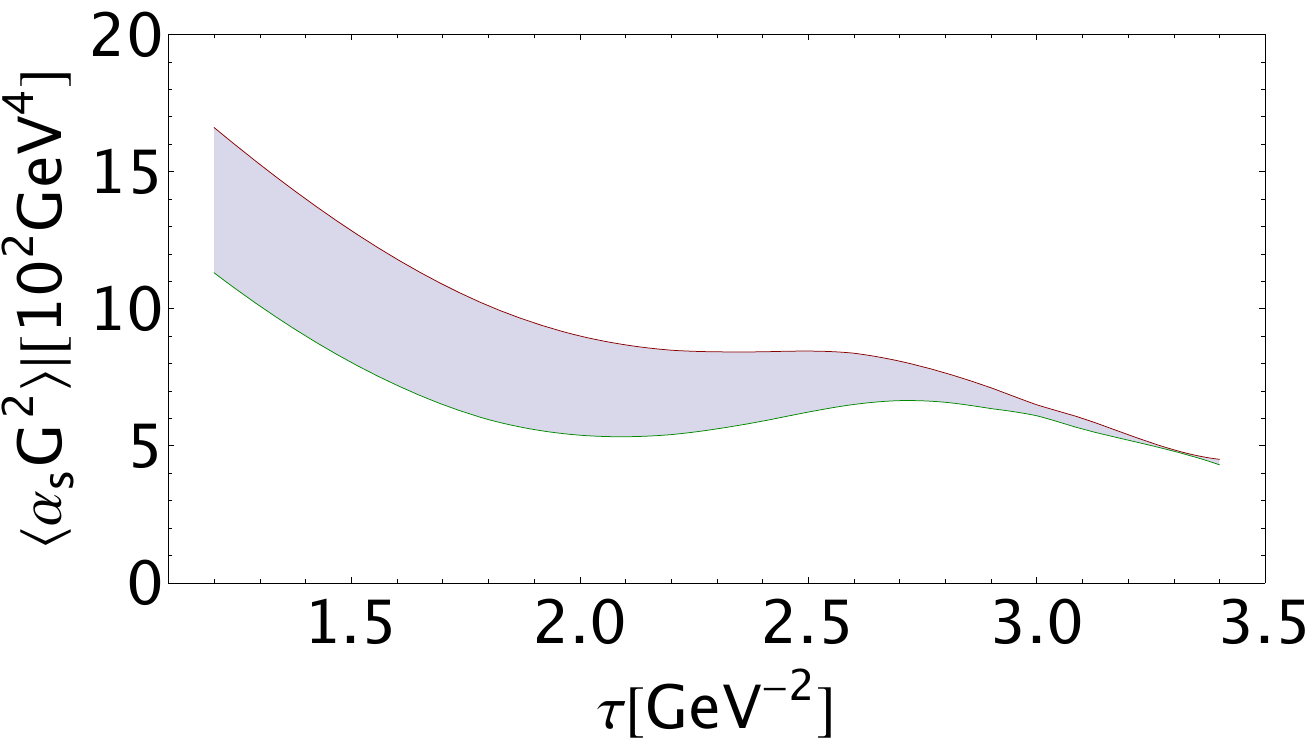}
\caption{\footnotesize  $\la \alpha_s G^2\ra$ versus the LSR variable $\tau$. } \label{fig:g2}
\end{center}
\vspace*{-0.5cm}
\end{figure} 
\subsection*{\b   $\la \alpha_s G^2\ra$ from $ {\cal R}^A_{10}$}  

We use the previous values of $d_{6,A}$ and $d_{8,A}$  into $ {\cal R}^A_{10}$ and we re-extract $\la \alpha_s G^2\ra$ using a one-parameter fit. The analysis is shown in Fig.\,\ref{fig:g2}. We obtain:
\beq
\la \alpha_s G^2\ra =(6.9\pm 1.5)\times 10^{-2}\,{\rm GeV}^4,
\eeq
in good agreement with the one in Eq.\,\ref{eq:g2} used previously as input. This result also indicates the sef-consistency of the set of condensates entering in the analysis. 
\section{BNP $\tau$-decay like moments}
As emphasized by BNP in Ref.\,\cite{BNP,BNP2}, it is more convenient to express the moment in terms of the combination of Spin (1+0) and Spin 0 spectral functions in order to avoid some eventual pole from $\Pi^{(0)}$ at $s=0$:
  \beq
  {\cal R}_{n,H}=6\pi\,i\,\int_{|s|=M_0^2} \hspace*{-0.5cm}dx\, (1-x)^2\,x^n\ga (1+2x) \,\Pi^{(1+0)}_H(x) -2x\Pi^{(0)}_H(x)\dr
  \eeq
  with $x\equiv s/M^2_0$, $H\equiv V,A$. $n$ indicates the degree of moment.  The lowest moment ${\cal R}_{0,A}$ corresponds to the physical $\tau$-decay process\,\cite{BNP2,BNP}. 
\subsection*{\b The lowest moment ${\cal R}_{0,A}$ from the data}  

\begin{figure}[hbt]
\begin{center}
\includegraphics[width=10cm]{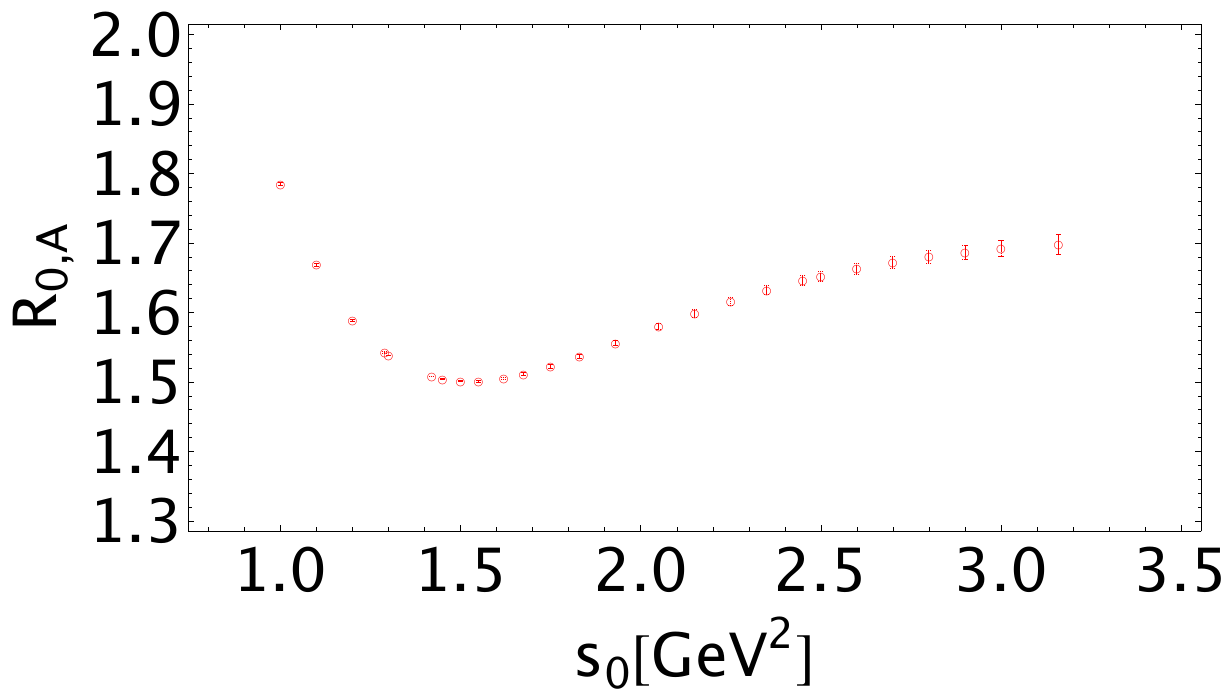}
\caption{\footnotesize  Lowest moment $R_{0,A}$ versus the hypothetical $\tau$-mass squared $s_0$. } \label{fig:r0A}
\end{center}
\vspace*{-0.5cm}
\end{figure} 
Using previous fits of the data, we show in Fig.\,\ref{fig:r0A} the behaviour of the lowest moment ${\cal R}_{0/A}$ versus an hypothetical $\tau$-lepton mass squared $s_0\equiv M^2_0$. At the observed mass value:  $M_\tau=1.777$ GeV, one obtains:
\beq
{\cal R}_{0,A}= 1.698(14).
\eeq
It reproduces with high accuracy the ALEPH value\,\cite{ALEPH} :  
\beq
{\cal R}_{0,A}\vert_{\rm Aleph}= 1.694(10),
\eeq

which is an important self-consistency test of our fitting procedure.

\subsection*{\b The lowest BNP moment ${\cal R}_{0,A}$}  
The QCD expression of ${\cal R}_{0,A}$ including the dimension six $(d_{6,A})$ and eight $(d_{8,A})$ condensates within the SVZ expansion\,\cite{SVZ} can be deduced from\,\cite{BNP}. To simplify for the reader, we give its expression, in the chiral limit\,\footnote{We use the complete expression including quark masses and $\alpha_s$ corrections in the numerical analysis.}:
\beq
{\cal R}_{0,A}= \frac{N_c}{2}|V_{ud}|^2\,S_{ew}\Big{\{} 1+\delta'_{ew}+\delta^{(0)}_A + \sum_{D=1,2,\dots}\delta_{A}^{(2D)}\Big{]}.
\eeq
where the electroweak factors and corrections  are :
\beq
V_{ud} = 0.97418,~~~~~~~~~~~~~~S_{ew}=1.019,~~~~~~~~~~~~~~\delta'_{ew}=0.0010.
\eeq
The QCD corrections copied from BNP are:

\d  Perturbative corrections to order $\alpha_s^4$:
\bea
 \delta^{(0)}_{A}\vert_{FO}&=&a_s+5.2023\,a_s^2+26.366\,a_s^3+ 127.079\,a_s^4,\nnb\\
 \delta^{(0)}_{A}\vert_{CI}&=&1.364\,a_s+2.54\,a_s^2+9.71\,a_s^3+ 64.29\,a_s^4,
\eea
for fixed order (FO)\,\cite{BNP} and contour improved (CI) \,\cite{LEDI}.

Observing that the PT series grows geometrically\,\cite{SNZ} from the calculated coefficients in different channels, we estimate the $a_s^5$ coefficient to be\,\cite{SNe} :
  \beq
 \delta_5^{FO} \approx \pm 552\,a_s^5, ~~~~~~~ ~~~~~~~~~~~~~~~\delta_5^{CI}\approx \pm 228\,a_s^5~,
  \label{eq:as5}
  \eeq
 which one can  consider either to be  the error due to the unknown higher order terms of the series or (more optimistically) to be the estimate of the uncalculated $\alpha_s^5$ coefficient. 
 
 \d  Power corrections up to $d_{8,A}$:
 
 They read:
  \bea
   \delta^{(2)}_{A}&=&-8(1+\frac{16}{3}a_s)\frac{(m_u^2+m_d^2)}{M_0^2}-4(1+\frac{25}{3}a_s)\frac{(m_um_d)}{M_0^2},\,\,\,\,\,\,\,\,\,\,\,\,\,\,\,\,\,\,\,\,
   \delta^{(2)}_{A}\vert_{tach}=-2\times 1.05\frac{\,a_s\lambda^2}{M_0^2},\nnb\\
   \delta^{(4)}_{A}&=&\frac{11\pi}{4}a_s^2\frac{\la\alpha_s G^2\ra}{M_0^4}
  +32\pi^2\ga1+\frac{63}{16}a_s^2\dr\frac{(m_u+m_d)\la\bar \psi_u\psi_u\ra}{M_0^4}
  -8\pi^2\sum_k \frac{m_k\la\bar \psi_k\psi_k\ra}{M_0^4} +{\cal O}(m_q^4) \nnb\\
   \delta^{(6)}_{A}&=& -6\frac{d_{6,A}}{M_0^6},\,\,\,\,\,\,\,\,\,\,\,\,\,\,\,\,\,\,\,\,\,\,\,\,\,\,\,\,\,\,\,\,\,\,\,\,\,\,\,\,\,
   \delta_{A}^{(8)}= -4\frac{d_{8,A}}{M_0^8},
  \eea
  where $s_0\equiv M_0^2$ while $d_{6,A}$ and $d_{8,A}$ have been defined in Eqs.\,\ref{eq:d8} . We have assumed $\la\bar \psi_u\psi_u\ra=\la\bar \psi_d\psi_d\ra$. We shall not include the $D=2$ contribution due to an eventual tachyonic gluon mass  within the standard OPE. Using duality\,\cite{SNZ}, this term can be included in the estimate of the non-calculated higher order terms of the PT series discussed previously. Instanton contributions are expected to have higher dimensions and their contributions can be safely neglected\,\cite{SNe}. 
\subsection*{\b $d_{6,A}$ and $d_{8,A}$ condensates from ${\cal R}_{0,A}$}  

\begin{figure}[hbt]
\begin{center}
\hspace*{-4cm} {\bf a)} \hspace*{8cm}{\bf b)}\\
\includegraphics[width=8cm]{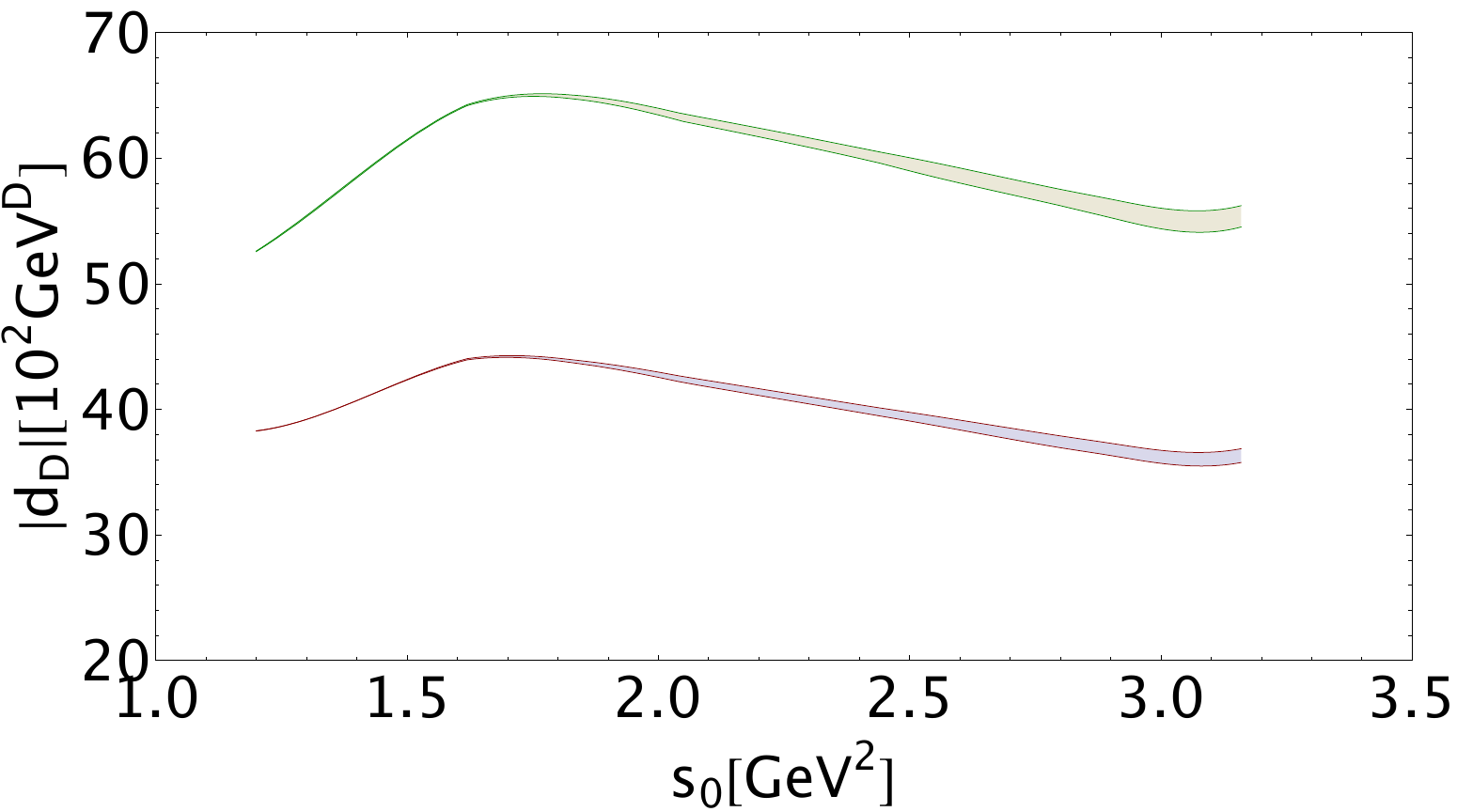}
\includegraphics[width=8cm]{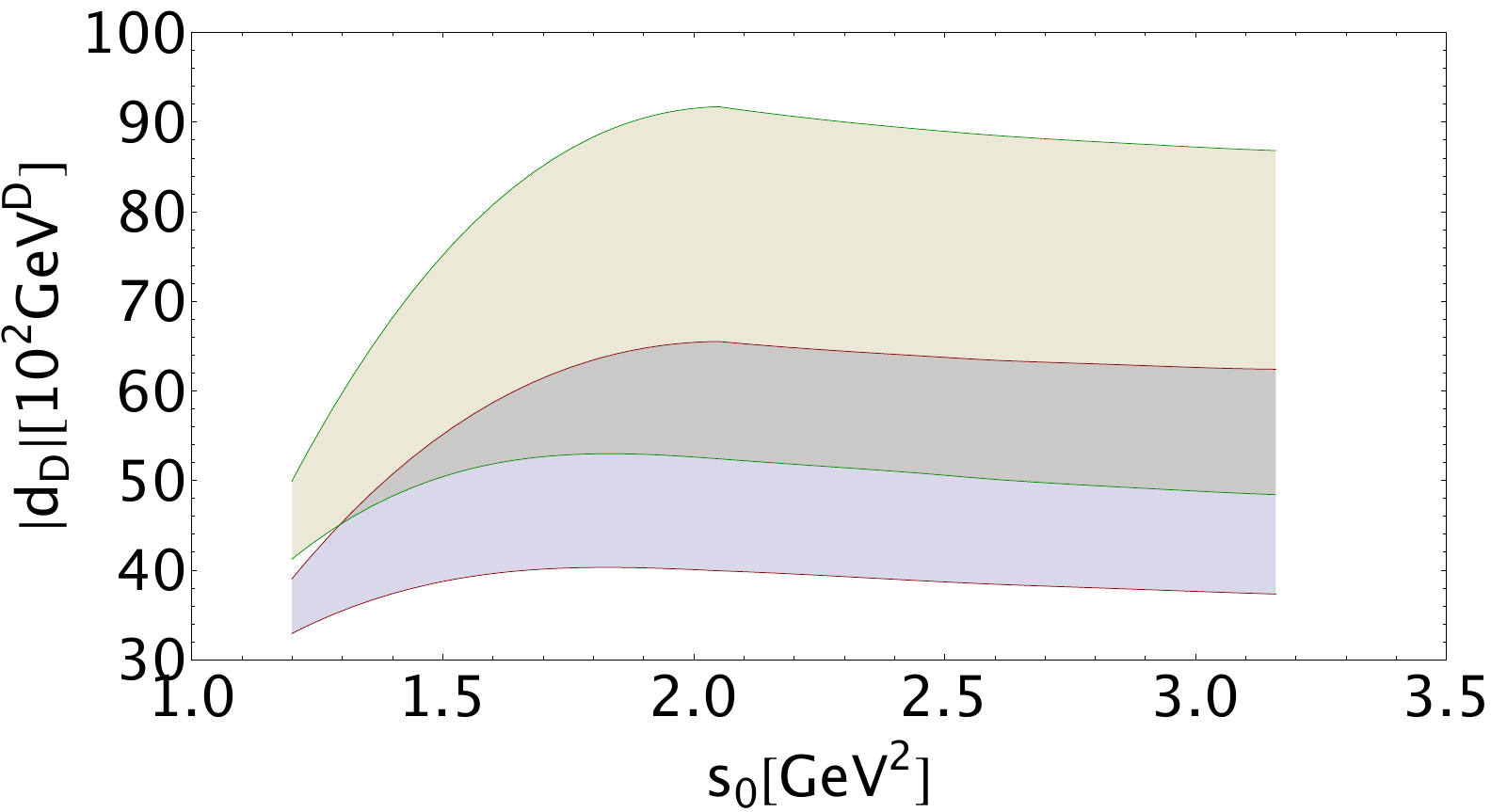}
\caption{\footnotesize  {\bf a)}: $|d_{6,A}|$ (lowest curve) and $|d_{8,A}|$ (highest curve)  versus  $M^2_0$; {\bf b)}: similar to  {\bf a)} but for $|d_{8,A}|$ and $|d_{10,A}|$. } \label{fig:d6-8}
\end{center}
\vspace*{-0.5cm}
\end{figure} 
We confront the experimental and QCD sides of ${\cal R}_{0,A}$ for different values of $s_0$. Like in the case of the ratio of Laplace sum rules,  we use a two-parameter fit ($d_{6,A},d_{8,A}$) to extract the values of these condensates using as input the values of the $d_4$ condensates and light quark masses. We consider the PT series up to order $\alpha_s^4$ and evaluate $\alpha_s$ at the hypothetical $\tau$-mass squared $s_0=M_0^2$. We use $\Lambda=(342\pm 8)$ MeV for $n_f=3$ flavours from the PDG world average\,\cite{PDG}. The results versus $s_0$ are shown in Fig.\,\ref{fig:d6-8}a). 

We consider as a reliable value the one from $s_0 = 1.93$ GeV$^2$ beyond the peak of the $A_1$ meson. We notice a stability (minimum) around 3 GeV$^2$ just below the physical $\tau$ mass which we consider as our optimal value:
\bea
d_{6,A} &=&  (36.2\pm 0.5)\times 10^{-2}\,{\rm GeV^6} ~ ~~~~~~~~~~~~~~ d_{8,A} = -(55.2\pm 0.9)\times 10^{-2}\,{\rm GeV^8} ~ ~~~~~~~~~~~~~~   {\rm (FO)} \nnb\\
d_{6,A}  &=&  (33.1\pm 0.5)\times 10^{-2}\,{\rm GeV^6} ~ ~~~~~~~~~~~~~~ d_{8,A} = -(50.6\pm 0.8)\times 10^{-2}\,{\rm GeV^8} ~ ~~~~~~~~~~~~~~   {\rm (CI)}
\eea
 One can notice from the analysis that the absolute values of the condensates are slightly higher for FO than for CI. To be conservative, we take the arithmetic average of the FO and CI values and add as a systematic the largest distance between the mean and the individual value\,:
\beq
d_{6,A} =  (34.6\pm1.8)\times 10^{-2}\,{\rm GeV^6} ~ ~~~~~~~~~~~~~~ d_{8,A} = -(52.9\pm2.4)\times 10^{-2}\,{\rm GeV^8}.
\label{eq:d68}
\eeq   

\section{ $d_{8,A}$ and $d_{10,A}$ from ${\cal R}_{1,A}$}  
The expression of ${\cal R}_{1,A}$ is similar to ${\cal R}_{1,V}$ given in Eq. 20 of Ref.\,\cite{SNe}.  Using a two-parameter fit  ($d_{8,A}, d_{10,A}$) of ${\cal R}_{1,A}$ for different $s_0$, we show the result
of the analysis in Fig.\,\ref{fig:d6-8}b) using FO  PT series.  One obtains:
\beq
d_{8,A} =  -(51.4\pm 11.0)\times 10^{-2}\,{\rm GeV^6} ~ ~~~~~~~~~~~~~~ d_{10,A} = (70.1\pm 21.6)\times 10^{-2}\,{\rm GeV^8}.
\label{eq:d810}
\eeq
 The results are stable versus $s_0$ but less accurate than in the case of ${\cal R}_{0,A}$ such that one cannot differentiate a FO from CI truncation of the PT series. Then, for higher moments, we shall only consider FO PT series.

\section{Final values of $d_{6,A}$ and $d_{8,A}$ } 
As a  final value of $d_{6,A}$, we take the mean of the values in Eqs.\,\ref{eq:d68-ratio} and \ref{eq:d68}. We obtain:
\beq
d_{6,A} = (34.4\pm 1.7)\times 10^{-2}\,{\rm GeV}^6,
\eeq
while for  $d_{8,A}$, we take the mean of the values obtained in Eqs. \,\ref{eq:d68-ratio}, \ref{eq:d68} and \ref{eq:d810}. We obtain:
\beq
d_{8,A} = -(51.51\pm 2.08)\times 10^{-2}\,{\rm GeV}^8.
\eeq
\d We notice that the relation\,:
\beq
d_{6,A}\simeq  -(11/7)\,d_{6,V},
\eeq
 is quite well satisfied within the errors.   This result also suggests a violation of the four-quark condensate vacuum saturation (see Eq.\,\ref{eq:d8})  similar to the one found from $e^+e^-\to$ Hadrons data\,\cite{SNe,SNe2}\,:
 \beq
 \rho\alpha_s\la\bar \psi\psi\ra^2 = (6.38\pm 0.30)\times 10^{-4}\,{\rm GeV}^6~~~~ \lrar2 ~~~~ \rho\simeq  (6.38\pm 0.30).
 \eeq
\subsection*{\b Determination of $\alpha_s(M_\tau)$}
\begin{figure}[hbt]
\begin{center}
\includegraphics[width=10cm]{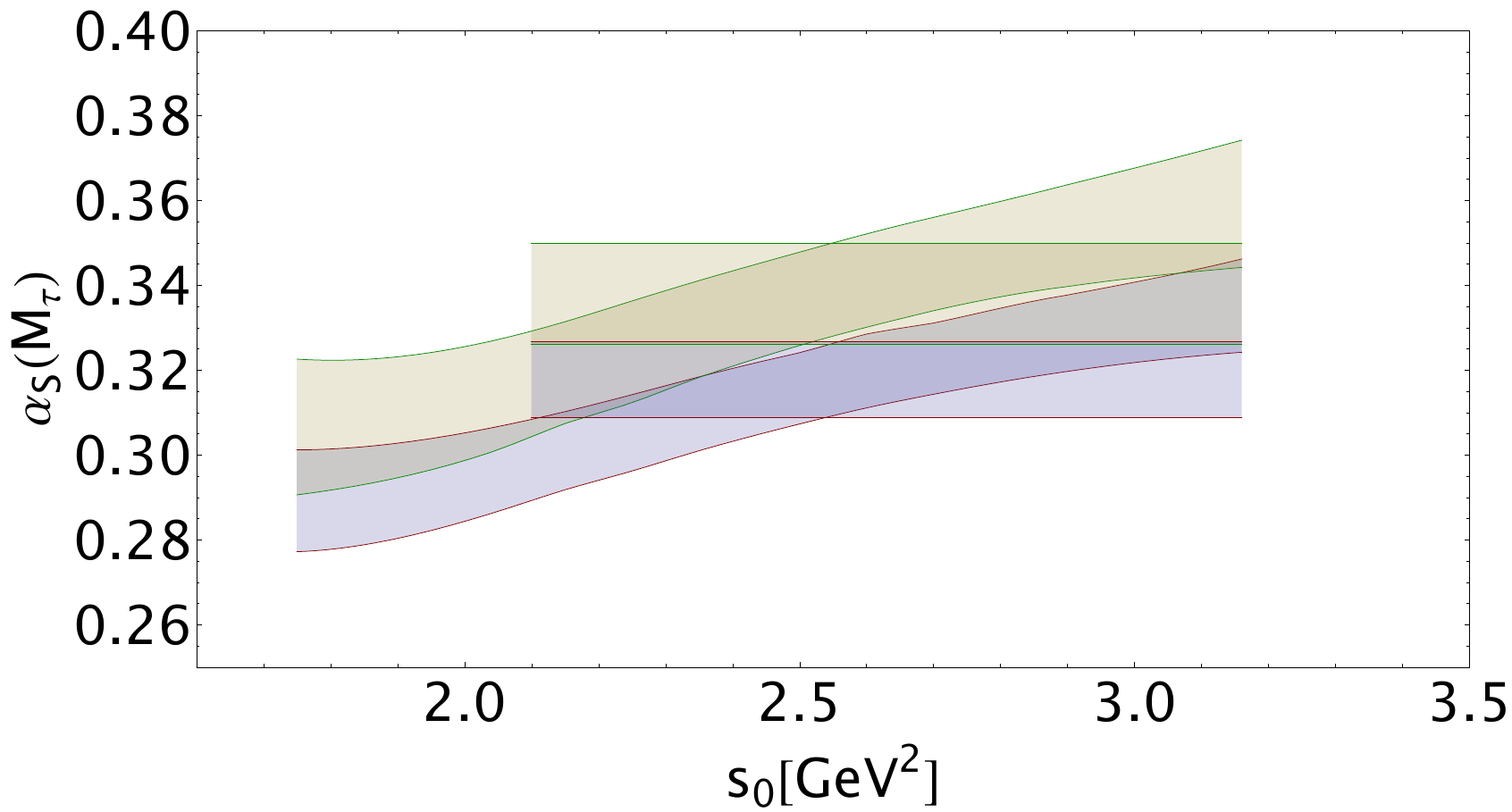}
\caption{\footnotesize  $\alpha_s(M_\tau)$  versus the hypothetical $\tau$-mass squared $s_0$. } \label{fig:as-A}
\end{center}
\vspace*{-0.5cm}
\end{figure}

We use the previous values of $d_{6,A} $ and $d_{8,A}$ together with the one of $\la\alpha_s G^2\ra$ in Eq.\ref{eq:g2} as inputs in the lowest BNP moment ${\cal R}_{0,A}$ in order to determine $\alpha_s(M_\tau)$. We show in Fig.\,\ref{fig:as-A}, the behaviour of $\alpha_s(M_\tau)$ versus an hypothetical $\tau$ mass squared $M_0^2\equiv s_0$. One can notice an inflexion point in the region $2.5^{+0.10}_{-0.15}$ GeV$^2$ at which we extract the optimal result. The conservative result
from $s_0=2.1$ GeV$^2$ to $M_\tau^2$ (see Fig.\,\ref{fig:as-A}) is\,:
\bea
\alpha_s(M_\tau)\vert_A &=&  0.3178(10) (65)~~~~~~~\lrar2~~~~~~~ \alpha_s(M_Z)\vert_A =  0.1182(8)(3)_{evol} ~ ~~~~~~~~~~~~~~   {\rm (FO)} \nnb\\
&=& 0.3380(10)(43)~ ~~~~~~~\lrar2~~~~~~~ \alpha_s(M_Z)\vert_A =  0.1206(5)(3)_{evol} ~ ~~~~~~~~~~~~~~   {\rm (CI)}.
\label{eq:as-A}
 \eea
 The 1st error in $\alpha_s(M_\tau)\vert_A$ comes from the fitting procedure. The  2nd one comes from  an estimate of the $\alpha_s^5$ contribution from Ref.\,\cite{SNe}.  At the scale $s_0=$2.5 GeV$^2$ the sum of non-perturbative contributions to the moment normalized to the parton model is:
 \beq
 \delta_{NP,A}\simeq -(7.9\pm 1.1) \times 10^{-2}.
 \eeq
One can notice that extracting $\alpha_s(M_\tau)\vert_A $ at the observed $\tau$-mass tends to overestimate the result:
 \beq
 \alpha_s(M_\tau)\vert_A = 0.3352 (40)~~~{\rm FO},  ~~~~~~~~~~ 0.3592(47)~~~{\rm CI}.
 \label{eq:as-tau-A}
 \eeq
 These values extracted at $M_\tau$ agree with the ones from Ref.\,\cite{PICH1} obtained at the same scale. The same feature has been observed in the case of $e^+e^-\to$ Hadrons. 

\subsection*{\b Comparison with previous results}
   {\scriptsize
   \begin{center}
\begin{table}[hbt]
\setlength{\tabcolsep}{1.1pc}
  \begin{center}
    {
  \begin{tabular}{ccc cc ll}

&\\
\hline
\hline
\oliva Channel  &\oliva$d_6$ &\oliva$-d_8$ &\oliva$\alpha_s(M_\tau)$ FO&\oliva$\alpha_s(M_\tau)$ CI &\oliva Refs.\\
 \hline 
$e^+e^-$&$-(26.3\pm 3.7)$&$-(18.2\pm 0.6)$&0.3081(86)&0.3260(78)& \cite{SNe}\\
V&&&0.3129(79)&0.3291(70) &\cite{SNe}\\ 
\hline
A&$34.4\pm 1.7$&$51.5\pm 2.1$&0.3157(65)&0.3368(45)&This work\,\\
 A&$43.4\pm 13.8$&$59.2\pm 19.7$&$0.3390(180)$&$0.3640(230)$&\,\cite{PICH1}\\
A & $19.7\pm 1.0$&$27\pm 1.2$&{--}&0.3350(120)& \,\cite{ALEPH} \\
A& $9.6\pm 3.3$&$9.0\pm 5.0$&0.3230(160)&0.3470(230) &\,\cite{OPAL} \\

   \hline\hline
\end{tabular}}
 \caption{Values of the QCD condensates from some other $\tau$-moments at Fixed Order (FO) PT series and of $\alpha_s(M_\tau)$ for FO and Contour Improved (CI) PT series.}\label{tab:other} 
 \end{center}
\end{table}
\end{center}
} 
We compare our results with the ones from $e^+e^-$ and $\tau$-decay Vector channel\,\cite{SNe} and with the results obtained by different
authors in the Axial-Vector channel:

\d Our values of  $d_{6,A}$ and $d_{8,A}$ are in good agreement within the errors with the ones of Ref.\,\cite{PICH1} but about two times larger than the ones of Ref.\cite{ALEPH}. 

\d The value of $d_{8,A}$ suggests that the assumption in Eq.\,\ref{eq:d8}:
\beq
d_{8,A}\approx d_{8,V}
\eeq
 is not satisfied by the fitted values given in Table\,\ref{tab:other}. 
\section{ High-dimension condensates }  
To determine the high-dimension condensates,  we  use the analogue of the moments given Eq. 19 of Ref.\,\cite{SNe} for the vector channel. 
\subsection*{\b $d_{10,A}$ and $d_{12,A}$ condensates from ${\cal R}_{2,A}$}  
\begin{figure}[hbt]
\begin{center}
\hspace*{-4cm} {\bf a)} \hspace*{8cm}{\bf b)}\\
\includegraphics[width=8cm]{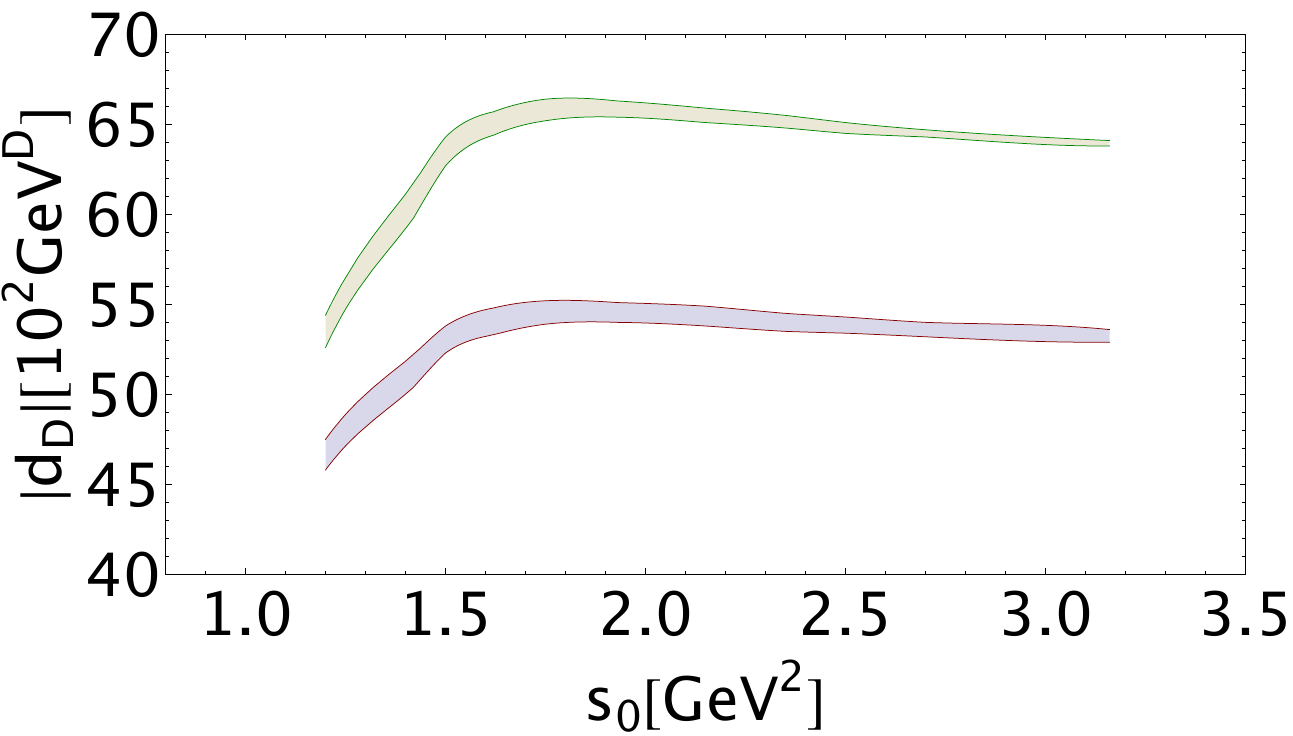}
\includegraphics[width=8cm]{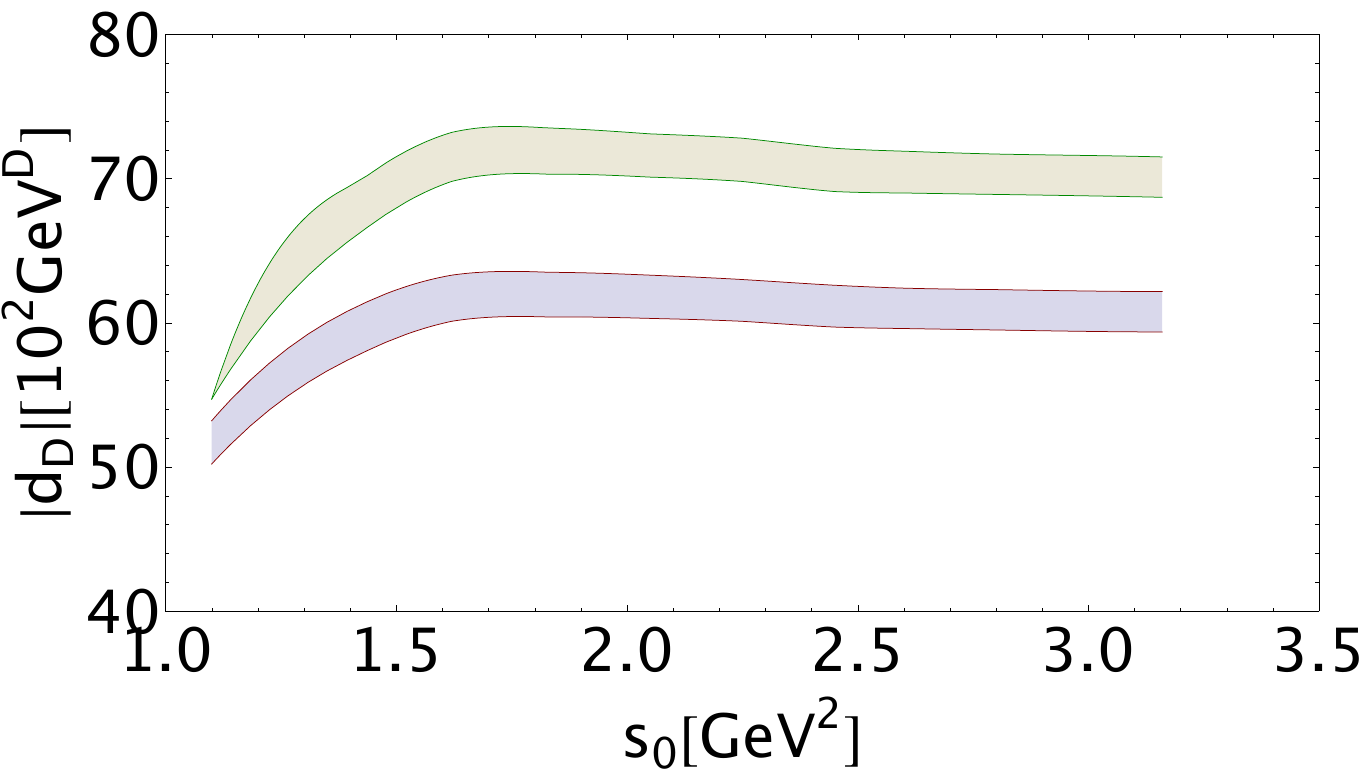}
\caption{\footnotesize  {\bf a)}: $|d_{10,A}|$ (lowest curve) and $|d_{12,A}|$ (highest curve)  versus  $M^2_0$; {\bf b)}: similar to  {\bf a)} but for $|d_{12,A}|$ and $|d_{14,A}|$. } \label{fig:d10-14}
\end{center}
\vspace*{-0.5cm}
\end{figure}

We  use a two-parameter fit to extract $(d_{10,A},d_{12,A})$ from    ${\cal R}_{2,A}$,  $(d_{12,A}, d_{14,A})$  from ${\cal R}_{3,A}$.
The $s_0$ behaviour of the results is given in Fig.\,\ref{fig:d10-14}.

We deduce from ${\cal R}_{2,A}$:
\beq
d_{10,A}=   (53.9\pm 0.7)\times 10^{-2}\,{\rm GeV}^{10},~~~~~~~~~~~~ 
d_{12,A}=  - (65.\pm 1.)\times 10^{-2}\,{\rm GeV}^{12}
\label{eq:1012}
\eeq
inside the stability region 1.93 GeV$^2$ to $M_\tau^2$. 
We take as a final value of $d_{10,A}$ the mean from ${\cal R}_{1,A}$ in Eq.\,\ref{eq:d810} and the one from ${\cal R}_{2,A}$ in Eq.\,\ref{eq:1012}:
\beq
d_{10,A}=   (53.9\pm 0.7)\times 10^{-2}\,{\rm GeV}^{10}.
\eeq
\subsection*{\b $d_{2n,A}$ and $d_{2(n+1),A}$ condensates}  
\begin{figure}[hbt]
\begin{center}
\hspace*{-4cm} {\bf a)} \hspace*{8cm}{\bf b)}\\
\includegraphics[width=8cm]{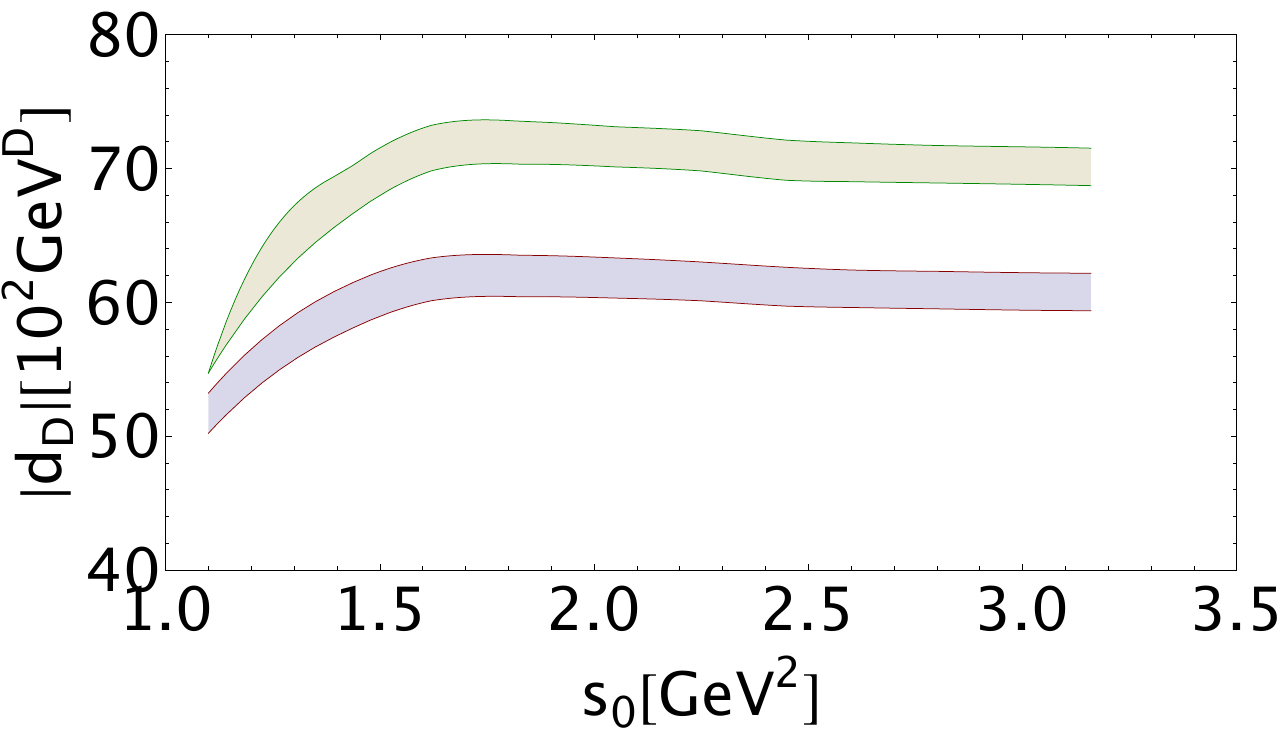}
\includegraphics[width=8cm]{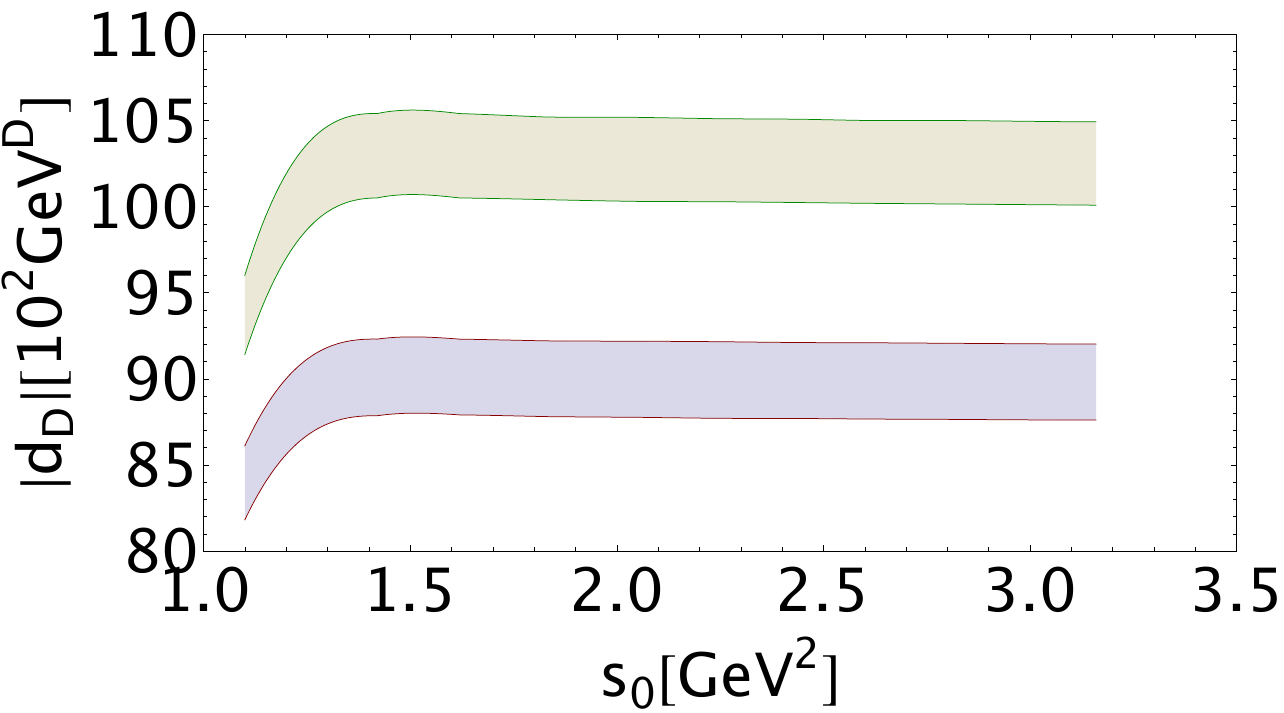}
\caption{\footnotesize  {\bf a)}: $|d_{14,A}|$ (lowest curve) and $|d_{16,A}|$ (highest curve)  versus  $M^2_0$; {\bf b)}: similar to  {\bf a)} but for $|d_{16,A}|$ and $|d_{18,A}|$. } \label{fig:d14-18}
\end{center}
\vspace*{-0.5cm}
\end{figure}

\begin{figure}[hbt]
\begin{center}
\includegraphics[width=10cm]{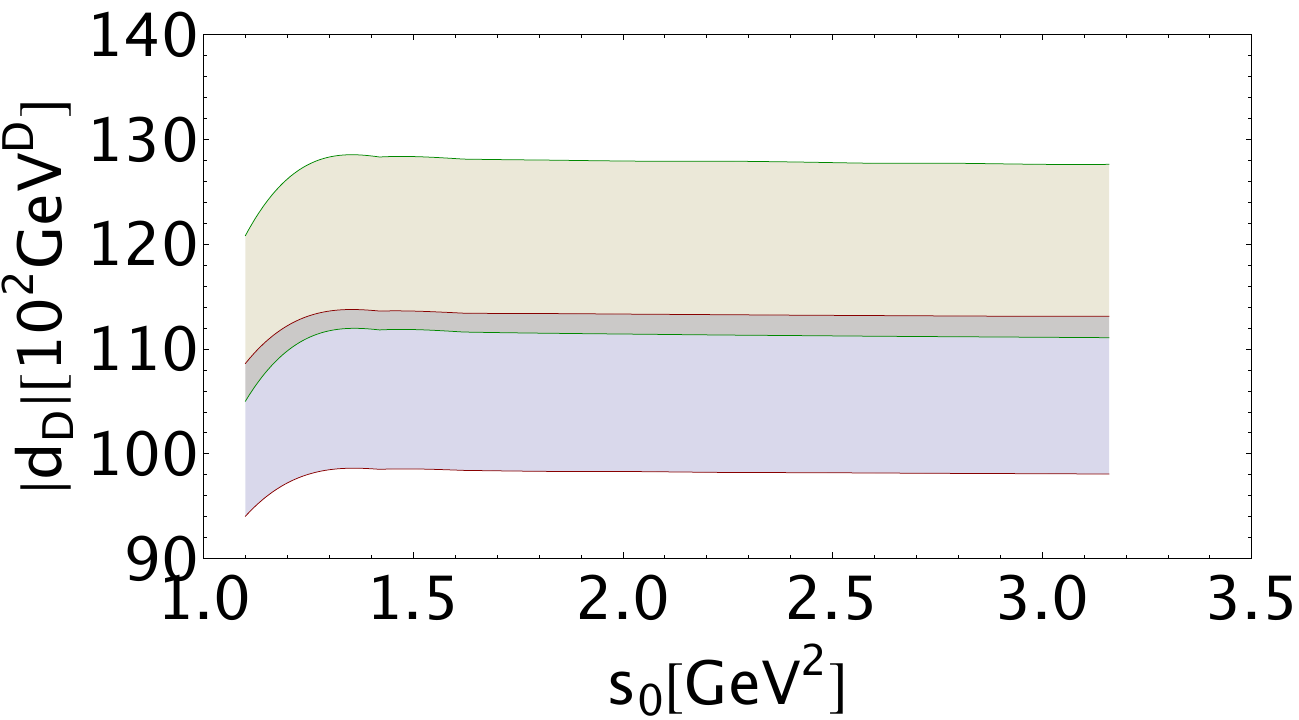}
\caption{\footnotesize  $|d_{18,A}|$ (lowest curve) and $|d_{12,A}|$ (highest curve)  versus  $M^2_0$. } \label{fig:d18-20}
\end{center}
\vspace*{-0.5cm}
\end{figure}

We do the same procedure as previously for higher dimension condensates $n\geq 6$.  The $s_0$  behaviours of the 
condensates are shown in Figs.\,\ref{fig:d14-18},\, \ref{fig:d18-20}. The results  are summarized in Table\,\ref{tab:cond-A}.

 \d One can notice that  the condensates in the A channel has alternate sign. 
 
 \d Their size is almost constant and more accurate than previous determinations in the literaure. There is not also any sign of an exponential growth. This feature in the Euclidian region does not favour  a sizeable duality violation in the time-like region\,\cite{SHIFMAN}. 

   {\normalsize
\begin{table}[H]
\setlength{\tabcolsep}{0.22pc}
  \begin{center}
    {
  \begin{tabular}{lllll ll ll}

&\\
\hline
\hline
$\oliva d_{6,A}$&$\oliva -d_{8,A}$&$\oliva d_{10,A}$&$\oliva -d_{12,A}$&$\oliva d_{14,A}$& $\oliva -d_{16,A}$&$\oliva d_{18,A}$&$\oliva- d_{20,A}$&Refs.\\
 \hline 
$34.4\pm 1.7$&$51.5\pm 2.1$&$53.9\pm 0.7$&$63.3\pm1.8$&$77.0\pm 6.0$&$93.1\pm 4.0$&$104.3\pm 7.7$&$ 119.7\pm 8.2$& This work\\
$43.4\pm 13.8$&$59.2\pm 19.7$&$63.2\pm 33.6$&$43.4\pm31.6$&&&&&\cite{PICH1}\\
 $19.7\pm 1.0$&$27\pm 1.2$&&&&&&&\cite{ALEPH} \\
 $9.6\pm 3.3$&$9.0\pm 5.0$&&&&&&&\cite{OPAL} \\
   \hline\hline
\end{tabular}}
 \caption{ Values of the QCD condensates of dimension $D$ in units of $10^{-2}$ GeV$^{D}$ from  this work and some other estimates. }\label{tab:cond-A} 
 \end{center}
\end{table}
} 
\section{The V--A channel}
\subsection*{\b The two-point function}
Its corresponds to the V--A quark current:
\beq
 J^\mu_{V-A}(x)=: \bar\psi_u\gamma^\mu(1-\gamma_5)\psi_d:.
\eeq
In the chiral limit, the QCD expression of the corresponding two-point function is similar to the one in Eq.\,\ref{eq:d8} where: 
\bea
d_{6,V-A} &=& \frac{1}{2}\ga d_{6,V}+d_{6,A}\dr 
= -\frac{1}{2}\ga \frac{4}{7} \dr d_{6,V} \nnb\\ &=& \frac{512}{27} \pi^3\rho\alpha_s\la\bar \psi\psi\ra^2 = (37.5\pm 1.9)\times 10^{-2}\,{\rm GeV}^{6},\nnb\\
d_{8,V-A} &=& \frac{1}{2}\ga d_{8,V}+d_{8,A}\dr \simeq -(14.5\pm 2.2)\times 10^{-2}\,{\rm GeV}^{8}
\eea
\subsection*{\b Data handling of the spectral function}

\begin{figure}[hbt]
\begin{center}
\includegraphics[width=10cm]{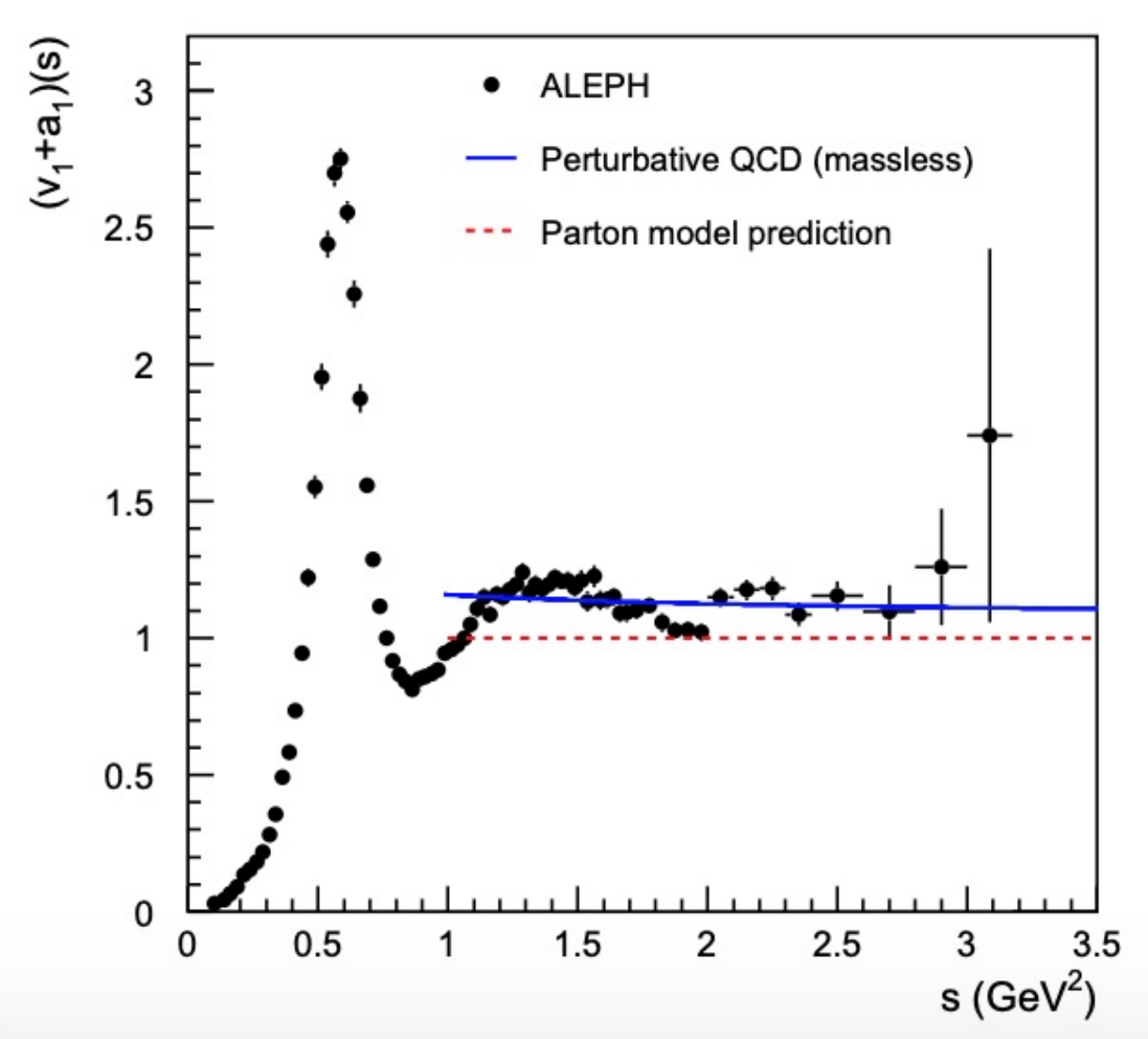}
\caption{\footnotesize  V--A spectral function measured by ALEPH. } \label{fig:V--A}
\end{center}
\vspace*{-0.5cm}
\end{figure}

\begin{figure}[H]
\begin{center}
\includegraphics[width=5.2cm]{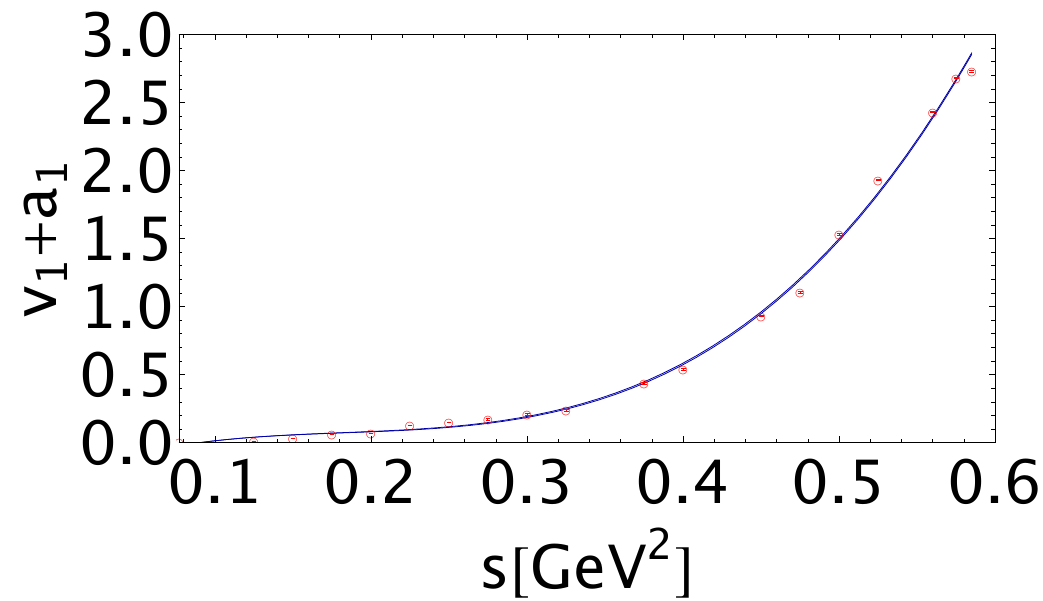}
\includegraphics[width=5.2cm]{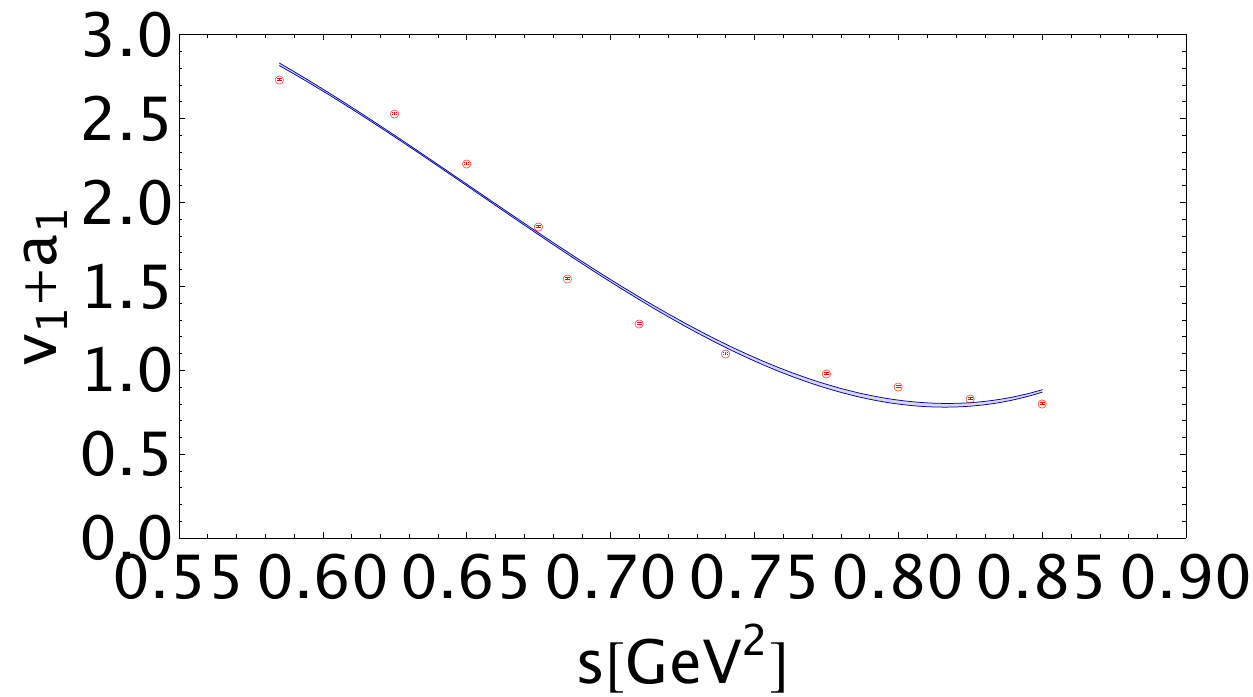}
\includegraphics[width=5.2cm]{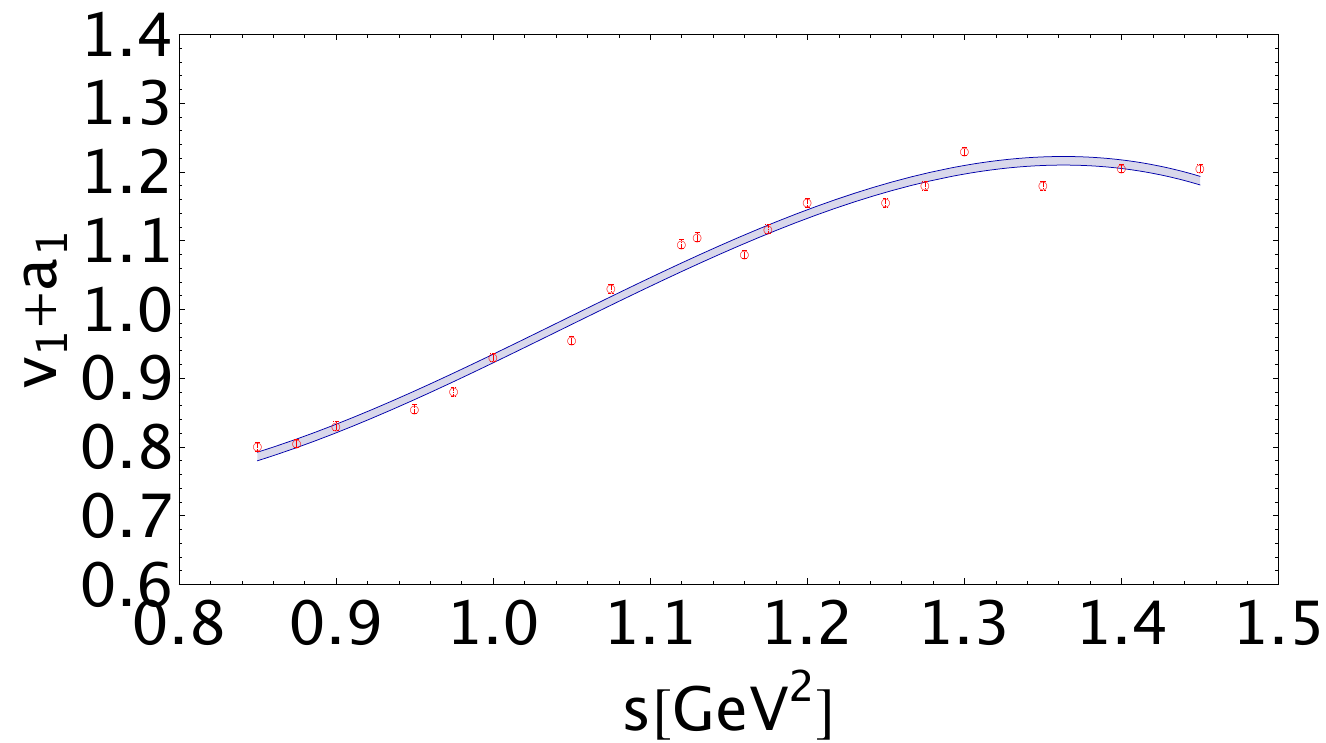}
\includegraphics[width=7cm]{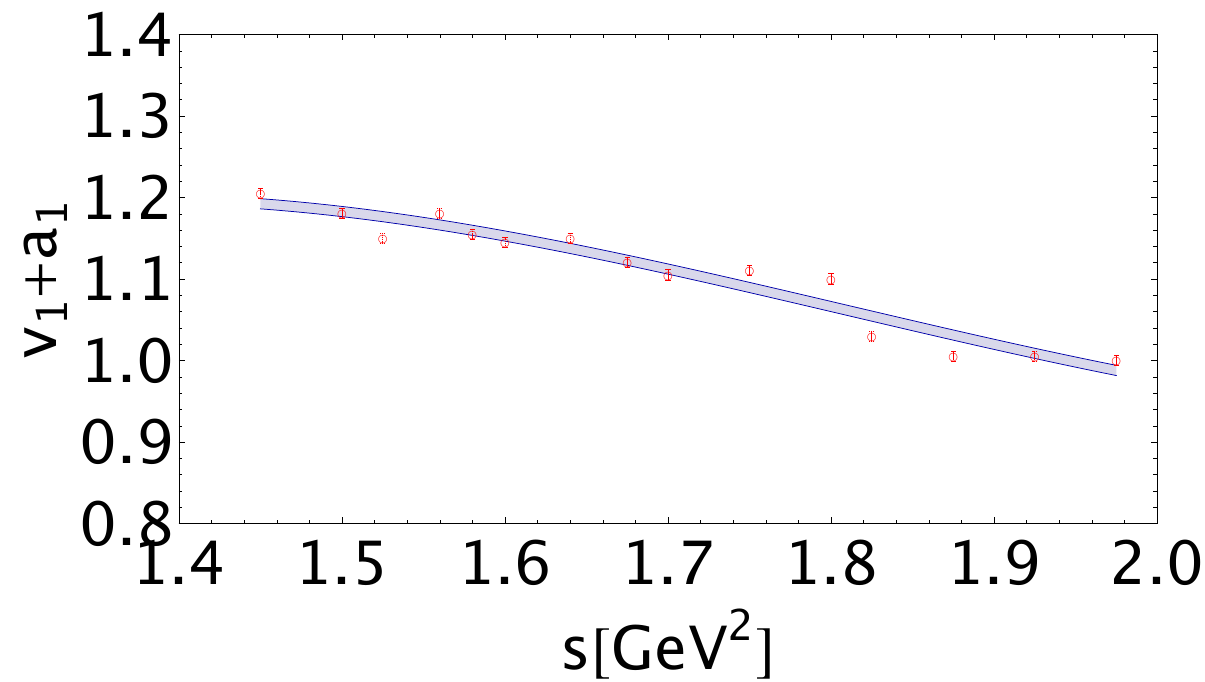}
\includegraphics[width=7.cm]{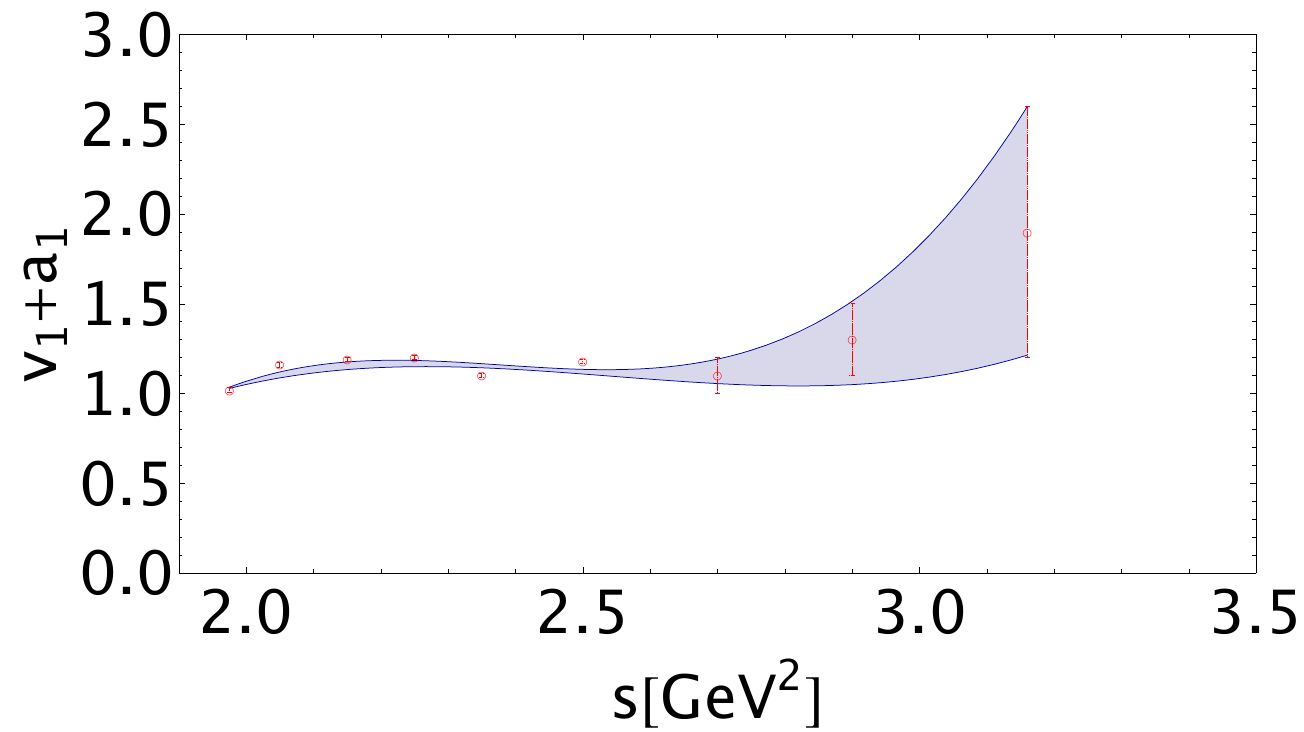}
\caption{\footnotesize  Fit of the data using a 2nd and 3rd order polynomial fits. } \label{fig:fitVA}
\end{center}
\vspace*{-0.5cm}
\end{figure}

The spectral function has been measured by ALEPH\,\cite{ALEPH} which we show in Fig.\.
,\ref{fig:V--A}.
In order to fit the data, we proceed like in the case of the axial-vector channel by subdividing the region into 5 subregions :
\beq
[4m_\pi^2,0.585],~ [0.585,0.85],~ [0.85,1.45],~ [1.45,1.975], ~[1.975, M_\tau^2],
\eeq
in units of GeV$^2$. We use  3rd order polynomials. The results of the fits are given in Fig.\,\ref{fig:fitVA}. 

\subsection*{\b The lowest BNP moment ${\cal R}_{0,V-A}$ } 
\begin{figure}[hbt]
\begin{center}
\includegraphics[width=10cm]{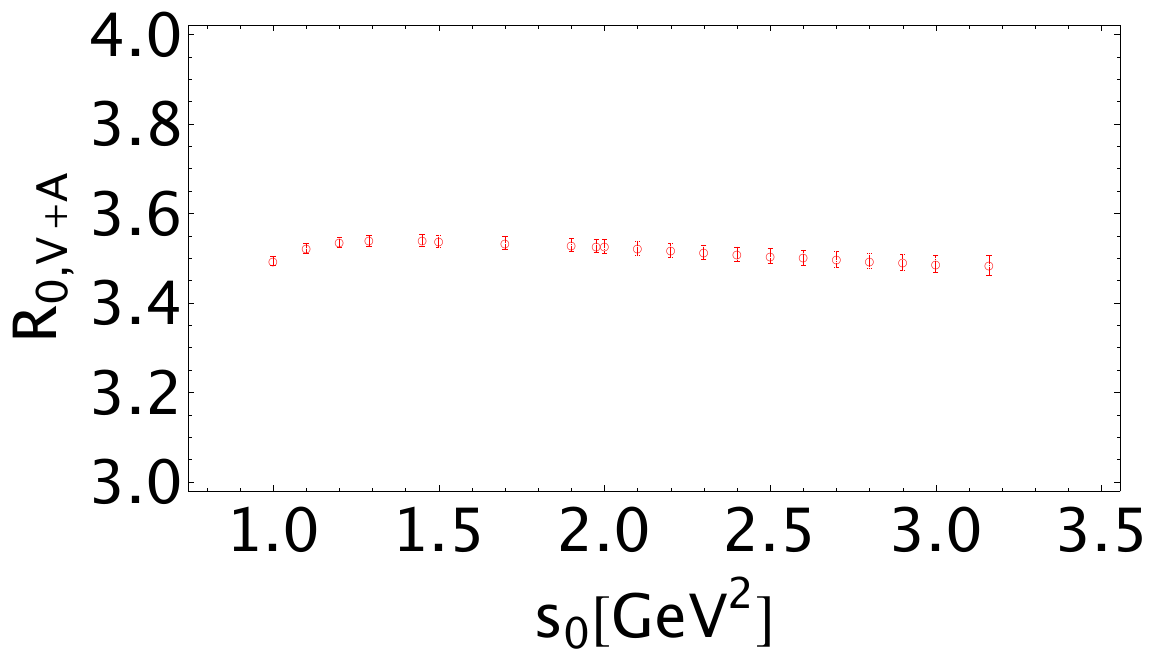}
\caption{\footnotesize  $s_0$ behaviour of the experimental moment ${\cal R}_{0,V--A}$.}
 \label{fig:mom-va}
\end{center}
\vspace*{-0.5cm}
\end{figure}

In the chiral limit their expression is similar to the previous ones for the axial-vector current modulo an overall factor 2 and the change into the condensates contributions $d_{D,V-A}$ of dimension $D$.
We show the $s_0$ behaviour of the moment in Fig.\ref{fig:mom-va}. 
For  $s_0=M_\tau^2$, we obtain :
\beq
{\cal R}_{0,V-A}= 3.484 \pm 0.022
\eeq
to be compared with the ALEPH data\,\cite{ALEPH}:
\beq
{\cal R}_{0,V-A}\vert_{Aleph} = 3.475\pm 0.011,
\eeq
where our error is larger which may  due to the fact that  we have  separately fitted the upper and lower values of the data. 
\subsection*{\b The $d_{6,V-A}$ and $d_{8,V-A}$ condensates}
We shall use the values of the corresponding  condensates from the $e^+e^-\to$ Hadrons
in Ref.\,\cite{SNe,SNe2} and the ones from the axial-vector channel  obtained in the previous section. They read:
\bea
d_{6,V-A} &\equiv& \frac{1}{2}\ga d_{6,V}+d_{6,A}\dr= +(3.6\pm 0.5)\times 10^{-2}\,{\rm GeV}^{6},\nnb\\
d_{8,V-A} &\equiv& \frac{1}{2}\ga d_{8,V}+d_{8,A}\dr= -(14.5\pm 2.2)\times 10^{-2}\,{\rm GeV}^{8},
\label{eq:d68-VA}
\eea
where the error is the largest relative \% error from each channel.  We note (as can be found in Ref.\,\cite{BNP}) that for ${\cal R}_{0,V-A}(s_0)$, the contribution of  $\la\alpha_s G^2\ra$ is $\alpha_s$ suppressed while the one of $d_{6,V-A}$ is smaller than in the individual V and A channels. Then, we expect that V--A is a {\it golden channel} for extracting $\alpha_s$. 
\subsection*{\b $\alpha_s$ from ${\cal R}_{0,V-A}(s_0)$}
\begin{figure}[hbt]
\begin{center}
\includegraphics[width=10cm]{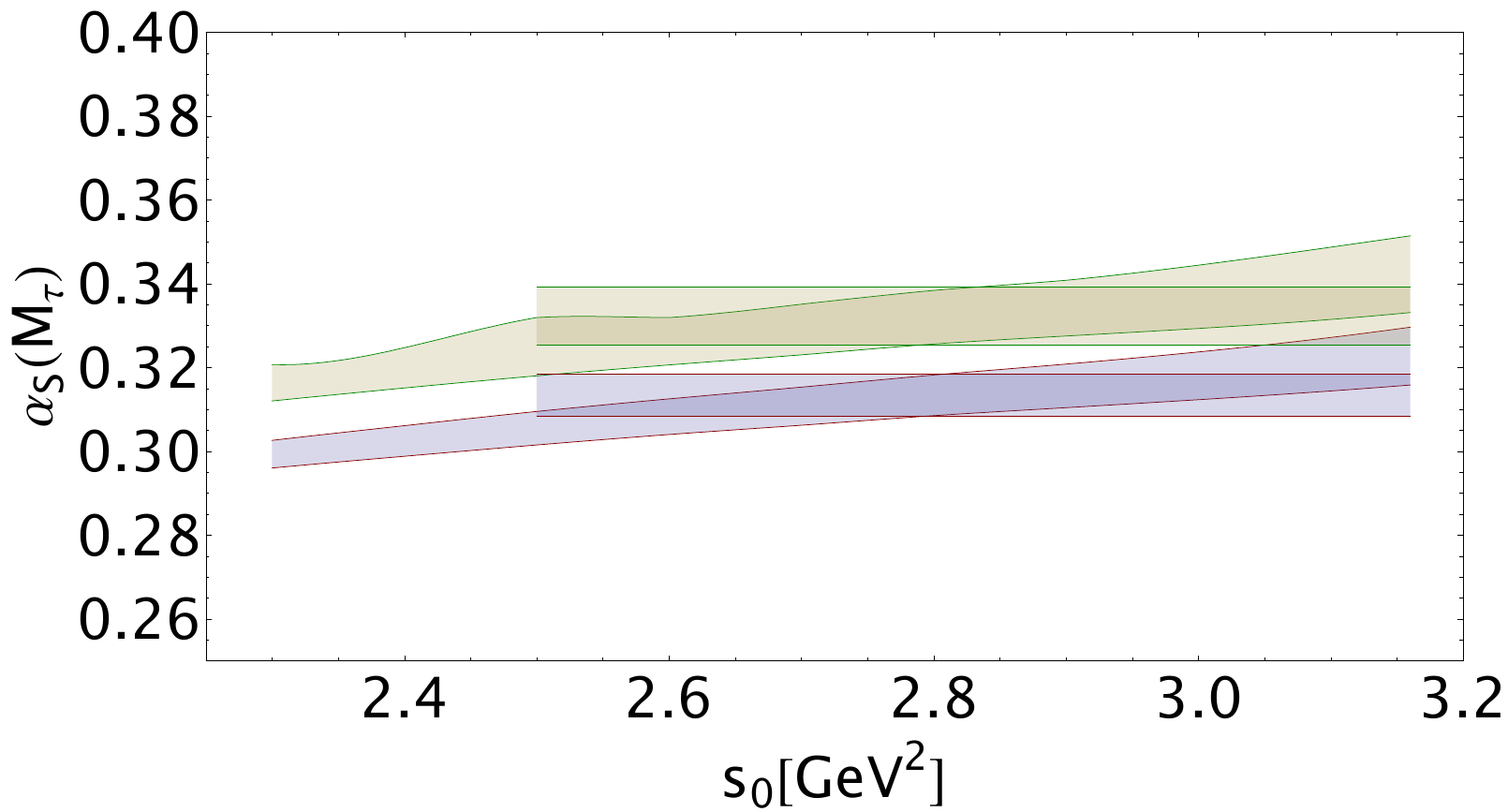}
\caption{\footnotesize  $\alpha_s(M_\tau)$  versus an hypothetical $\tau$-mass squared $s_0$. The upper curves corresponds to CI perturbative series and the lower ones to FO. The horizontal lines come from a least-square fit of the data in the optimal region $s_0\simeq (2.3\sim2.9)$ GeV$^2$.} \label{fig:as-VA}
\end{center}
\vspace*{-0.5cm}
\end{figure} 

Using the value of $\la\alpha_s G^2\ra$ in Eq.\,\ref{eq:g2} and the previous values of $d_{6,V-A}$ and $d_{8,V-A}$ in Eq,\,\ref{eq:d68-VA}, we extract the value of $\alpha_s(M_\tau)$ as a function of $s_0$ (see Fig.\,\ref{fig:as-VA}) from ${\cal R}_{0,V-A}(s_0)$. We notice a stable result in the region $s_0\simeq (2.5\sim 2.6)$ GeV$^2$ though not quite convincing. The, we consider as a conservative value the one obtained from a least-square fit of the values inside the region  $[2,5,M_\tau^2]$. The optimal result corresponds $s_0=2.8$ GeV$^2$ (see Fig.\,\ref{fig:as-VA}):
\bea
\alpha_s(M_\tau)\vert_{V-A} &=&  0.3135(51) (65)~~~~~~~\lrar2~~~~~~~ \alpha_s(M_Z)\vert_{V-A} =  0.1177(10)(3)_{evol} ~ ~~~~~~~~~~~~~~   {\rm (FO)} \nnb\\
&=& 0.3322(69)(43)~ ~~~~~~~\lrar2~~~~~~~ \alpha_s(M_Z)\vert_{V-A} =  0.1200(9)(3)_{evol} ~ ~~~~~~~~~~~~~~   {\rm (CI)}.
\label{eq:as-VA}
 \eea
 
 The 1st error in $\alpha_s(M_\tau)\vert_{V-A}$ comes from the fitting procedure. The 2nd one from  an estimate of the $\alpha_s^5$ contribution from Ref.\,\cite{SNe}.  At this scale the sum of non-perturbative contributions to the moment normalized to the parton model is:
 \beq
 \delta_{NP,V-A}\simeq +(2.7\pm 1.1) \times 10^{-4},
 \eeq
 which is completely negligible.
One can notice from Fig.\,\ref{fig:as-VA} that extracting $\alpha_s(M_\tau)\vert_{V-A} $ at the observed $M_\tau$-mass tends to overestimate its value\,:
 \beq
 \alpha_s(M_\tau)\vert_{V-A}= 0.3227 (69)(65)~~~{\rm FO},  ~~~~~~~~~~ 0.3423(92)(43)~~~{\rm CI}.
 \label{eq:as-tau-VA}
 \eeq

\subsection*{\b Comparison with some previous results}
   {\scriptsize
   \begin{center}
\begin{table}[hbt]
\setlength{\tabcolsep}{1.1pc}
  \begin{center}
    {
  \begin{tabular}{ccc cc ll}

&\\
\hline
\hline
\oliva$d_{6,V-A}$ &\oliva$-d_{8,V-A}$&\oliva$\alpha_s(M_\tau)$ FO&\oliva$\alpha_s(M_\tau)$ CI  & \oliva $s_0$ [GeV]$^2$&\oliva Refs.\\
 \hline 
$3.6\pm 0.5$&$14.5\pm 2.2$&0.3135(83)&0.3322(81)& 2.5 $\to M_\tau^2$&This work\,\\
\hline
$3.6\pm 0.5$&$14.5\pm 2.2$&0.3227(95)&0.3423(102)& $M_\tau^2$&This work\,\\
 $5.1^{+5.5}_{-3.1}$&$3.2\pm 2.2$&$0.3170^{+0.0100}_{-0.0050}$&$ 0.3360^{+0.0110}_{- 0.0090}$&$M_\tau^2\,(D\leq10)$ &Table 7 \,\cite{PICH1}\\
  $24^{+24}_{-16}$&$32^{+25}_{-32}$&$0.3290^{+0.0120}_{-0.0110}$&$ 0.3490^{+0.0160}_{- 0.0140}$&$M_\tau^2\,(D\leq12)$ &Table 8 \,\cite{PICH1} \\

  $2.4\pm 0.8$&$3.2\pm 0.8$&--&$0.3410(78)$&$M_\tau^2$&\,\cite{ALEPH}\\
    $1.5\pm 4.8$&$3.7\pm 9.0$&0.3240(145)&$0.3480(212)$&$M_\tau^2$&\,\cite{OPAL}\\
    
   \hline\hline
\end{tabular}}
 \caption{Values of the QCD condensates from ${\cal R}_{0,e^+e^-}$ and ${\cal R}_{0,A}$ at Fixed Order (FO) PT series and of $\alpha_s(M_\tau)$ for FO and Contour Improved (CI) PT series. The condensates are in units of $10^{-2}$ GeV$^D$.}\label{tab:other-VA} 
 \end{center}
\end{table}
\end{center}
} 
In Table\,\ref{tab:other-VA}, we compare our results with some other determinations\,\cite{ALEPH,PICH1,OPAL}\,\footnote{Some estimates including renormalon within a large $\beta$ approximation [ resp. duality violation] effects can be e.g. found  in Refs.\,\cite{CVETIC}  [resp. \cite{DV}].}  obtained at the scale $s_0=M_\tau^2$.  

\d There is a quite good agreement for $d_{6,V-A}$ but not for $d_{8,V-A}$ from different determinations.

\d The central values of the condensates given in  Table 7 of Ref.\,\cite{PICH1} using a truncation of the OPE up to $D=10$ are systematically smaller  than the ones in their Table 8 using the OPE up to $D=12$ though they agree within the errors.

\d Extracting  $\alpha_s$ at $M_\tau$, there is a good agreement among different determinations where the values are slightly higher than the ones from the optimal region  given in the first row\, (see Fig.\,\ref{fig:as-VA}). 
\section{Mean value of $\alpha_s$ from $e^+e^-\to$ Hadrons and A, V--A  $\tau$-decays}  
Using the result from $e^+e^-\to$ Hadrons and from the A and V--A  $\tau$-decay channels, we deduce the mean:
\bea
 \alpha_s(M_\tau)&=&0.3140(44)\, {\rm (FO)} ~ ~~~~~~~\lrar2~~~~~~~ \alpha_s(M_Z) = 1178(6)_{fit}(3)_{evol.},\nnb\\
 &=& 0.3346 (35) \, {\rm (CI)} ~ ~~~~~~~\lrar2~~~~~~~\alpha_s(M_Z)=0.1202(4)_{fit}(3)_{evol.}
 \eea
\section{Summary}  
\b We have improved the determinations of the QCD condensates in the axial-vector channel using a  ratio of LSR and higher BNP-like moments.   Our results are summarized in Table\,\ref{tab:cond-A}. We observe alternate signs, an almost constant value of their size. The absence of an exponential behaviour in the Euclidian region may not favour a duality violation in the time-like region\,\cite{SHIFMAN}. 

\b We have used as inputs the value of $\la\alpha_s G^2\ra$ better determinaed from the heavy quark channels and the ones of  the previous condensates  $d_{6,A}$ and $d_{8,A}$ to extract $\alpha_s(M_\tau)$ from the lowest BNP-moment ${\cal R}_{0,A} (s_0)$. Our conservative result in Eq.\,\ref{eq:as-A} is obtained from $s_0\simeq 2.1$ GeV$^2$ to $M_\tau^2$ where we notice from Fig.\,\ref{fig:as-A} that extracting $\alpha_s(M_\tau)$ at the observed $\tau$-mass leads to an overestimate of its value.

\b We combine the previous values of the $d_{6,A}$ and $d_{8,A}$ with the ones of the $d_{6,V}$ and $d_{8,V}$ from $e^+e^-\to$ Hadrons\,\cite{SNe,SNe2} into the lowest moment ${\cal R}_{0,V-A} (s_0)$ in the V--A channel.  Then, we extract the conservative value of  $\alpha_s(M_\tau)$ given in Eq.\,\ref{eq:as-tau-VA} at $s_0=[2.5$ GeV$^2\to M_\tau^2$]. Like in the case of the $e^+e^-\to$ Hadrons and axial-vector channel, we also notice that extracting $\alpha_s(M_\tau)$ at $M_\tau$ leads to an overestimate. 

\b We have not considered some eventual effects beyond the SVZ-expansion as:

\hspace*{0.5cm} \d According to Ref.\,\cite{SNZ}, the effect a tachyonic gluon mass\,\cite{CNZ} for  parametrizing phenomenologically the UV renormalon effect is equivalent by duality to the contribution of the uncalculated higher order PT terms. We assume that these terms  are well approximated by the estimate of the $\alpha_s^5$ term done in the paper using the observation that the coefficients of the calculated PT terms behave as a geometric sum.  

\hspace*{0.5cm} \d The observation that the calculated PT terms behave as a geometric sum and that no signal of alternate sign of these PT terms do not (a priori) favour the motivation  of a large $\beta$-approximation for the estimate of the UV renormalon.

\hspace*{0.5cm} \d The non-observation of an exponential behaviour of the non-perturbative condensate effects in the Euclidian region may indicate that Duality Violation in the time-like region\,\cite{DV} may not be sizeable\,\cite{SHIFMAN}.

\hspace*{0.5cm} \d Instanton effects act as high-dimension operators and their effects have been shown to be negligible in the vector channel\,\cite{SNe}. We expect similar features for the A and V--A channels.
\section*{Acknowledgement}
I thank Toni Pich for a careful reading of the manuscript and the 2nd referee for some constructive comments.
 \vspace*{-0.25cm}

\end{document}